\documentclass[structabstract]{aa}  
\usepackage{natbib}
\usepackage{graphicx}
\usepackage{longtable}
\usepackage{rotating}
\usepackage{lscape}

\usepackage{txfonts}
\begin{document}

 \title{A spectroscopy study of nearby late-type stars, possible members of stellar kinematic groups
 \thanks{
 Based on observations collected at the Centro Astron\'omico Hispano
 Alem\'an (CAHA) at Calar Alto, operated jointly by the Max-Planck Institut
 f\"{u}r Astronomie and the Instituto de Astrof\'isica de Andaluc\'ia (CSIC) and
 observations made with the Italian Telescopio Nazionale Galileo (TNG) operated
 on the island of La Palma by the Fundaci\'on Galileo Galilei of the INAF 
 (Istituto Nazionale di Astrofisica) at the Spanish Observatorio del Roque de 
 los Muchachos of the Instituto de Astrof\'isica de Canarias}
 \thanks{Appendices and Tables 5-15 are only available in the electronic version of the paper.  
 Table 1 is also available at the CDS via anonymous ftp to cdsarc.u-strasbg.fr (130.79.128.5)
 or via http://cdsweb.u-strasbg.fr/cgi-bin/qcat?J/A+A/
 }}


    \author{J. Maldonado
          \inst{1}
          \and
          R.M. Mart\'inez-Arn\'aiz \inst{2} \and C. Eiroa \inst{1} 
          \and D. Montes \inst{2}  \and B. Montesinos \inst{3}
          }

   \institute{Universidad Aut\'onoma de Madrid, Dpto. F\'isica Te\'orica, M\'odulo 15,
             Facultad de Ciencias, Campus de Cantoblanco, E-28049 Madrid, Spain,
             \and
             Universidad Complutense de Madrid, Dpto. Astrof\'isica, Facultad Ciencias F\'isicas,
             E-28040 Madrid, Spain 
             \and
             Laboratorio de Astrof\'isica Estelar y Exoplanetas, Centro de Astrobiolog\'ia, LAEX-CAB (CSIC-INTA),
             ESAC Campus, P.O. BOX 78, E-28691, Villanueva de la Ca\~{n}ada, Madrid, Spain   
             }
 
   \offprints{J. Maldonado, \\  \email{jesus.maldonado@uam.es}}
              
   \date{Received ; Accepted}

 
\abstract
 {Nearby late-type stars are excellent targets for seeking young objects
 in stellar associations and moving groups. 
 The origin of these structures is still misunderstood, and lists of moving group
 members often change with time and also from author to author. 
 Most members of these groups have been identified by means of kinematic criteria, leading to
 an important contamination of previous lists by old field stars. 
 }
 {We attempt to identify unambiguous moving group members among a sample
 of nearby-late type stars by studying their kinematics, lithium abundance, chromospheric
 activity, and other age-related properties.}
 {High-resolution echelle spectra ($R \sim 57000$) of a sample of nearby late-type stars are used
  to derive accurate radial velocities that are combined with the precise Hipparcos parallaxes
  and proper motions to compute galactic-spatial velocity components. Stars are classified as possible
  members of the classical moving groups according to their kinematics.
  The spectra are also used to study several age-related properties  for 
  young late-type stars, i.e., the equivalent width of the lithium  Li~{\sc i} \space 6707.8 \space \AA \space line or
  the  $R'_{\rm HK}$ index. 
  Additional information like X-ray fluxes from the ROSAT All-Sky Survey or the presence of debris
  discs is also taken into account. The different age estimators are compared and the moving group
  membership of the kinematically selected candidates are discussed.  
  }
  {From a total list of 405 nearby stars, 102 have been classified as moving group candidates
  according to their kinematics.  i.e., only  $\sim$ 25.2 \% of the sample. The number reduces when
  age estimates are considered, and only 26 moving group candidates (25.5\% of the 102 candidates) have ages
  in agreement with the star  having the same age as an MG member.}
   {}


   \maketitle


\section{Introduction}
\label{introduction} 
 
 Last years have been very productive in identifying small associations and 
 kinematic groups of young late-type stars in the solar vicinity. Although the study of
 moving groups (MGs) goes back more than one century, their origin and evolution remain still
 unclear,  
 and this term is commonly used in the literature to indicate any system of 
 stars sharing a common spatial motion. 
 The best-studied MGs are the so-called \textit{classical} MGs.
 Examples are Castor, IC2391, Ursa Major, the Local Association and the Hyades
 (e.g. Montes et al. 2001b; L\'opez-Santiago et al. 2006, 2009, 2010, and references therein).
 \nocite{2001MNRAS.328...45M,Javi06,2009A&A...499..129L,2010arXiv1002.1663L}

 In the classical theory of MGs developed by O. Eggen \citep{Eggen1994}, 
 moving groups are the \textit{missing link} between
 stars in open clusters and associations on one hand and field stars on the other. Open
 clusters are disrupted by the gravitational interaction with massive objects in the Galaxy
 (like giant molecular clouds), and as a result, the open cluster members are stretched out into
 a ``tube-like'' structure and dissolve after several galactic orbits. The result of 
 the stretching is that the stars appear, if the Sun happens to be inside the ``tube'', all
 over the sky, but they may be identified as a group through their common space velocity.

 Clusters disperse on time scales of a few hundred years  \citep{1971A&A....13..309W};
 therefore, most of these groups should be moderately young ($\sim$ 50 - 650 Myr). 
 However, Eggen's hypothesis is controversial and some of the MGs may also be the result of
 resonant dynamical structures.  
 For instance, \cite{2007A&A...461..957F}  studied a large sample of
 stars in the Hyades MG, and determined that it is a mixture of stars evaporated from the Hyades
 cluster and a group of older stars trapped at a resonance. MGs may also be produced by the
 dissolution of larger stellar aggregates, such as stellar complexes or fragments of
 old spiral arms.
 
 The \textit{young} MGs (8 - 50 Myr) are probably the most immediate dissipation products
 of the  youngest associations.
 Examples of such associations are TW Hya, $\beta$ Pic, AB Dor, $\eta$ Cha, $\epsilon$ Cha,
 Octans, Argus, the Great Austral complex (GAYA), and the Hercules-Lyra association
 \citep{Zuckerman,2008hsf2.book..757T,2004AN....325....3F,Javi06,David10}. 
 Some of the young MGs are in fact related to
 star-forming regions like the
 Scorpius-Centaurs-Lupus complex \citep{Zuckerman}, Ophiuchus or Corona Australis \citep{2007ApJS..169..105M}.\\
 
 The availability of  accurate parallaxes provided by the Hipparcos satellite became
 a milestone in the study of MGs. Statistical, unbiased studies of large samples of stars
 have confirmed the existence of the classical MGs and have given rise to new clues and theories about
 the origin of such structures. Examples of these studies are those by
 \cite{1999A&AS..135....5C}, \cite{1999A&A...341..427A}, \cite{1999MNRAS.308..731S}, and \cite{2008A&A...490..135A}.\\ 

 Identifying a star or group of stars as members of an MG is not a trivial task, and in fact, lists 
 of members change among different works.
 Most members of MGs have been identified by means of kinematic criteria; however, this is not sufficient 
 since many old stars
 can share the same spatial motion of those stars in MGs. For example, \cite{2009A&A...499..129L}
 show that among previous lists of Local Association members, roughly 30\%
 are old field stars.
 The membership issue can be partially solved if high-resolution spectroscopy is used.
 Recent studies have shown  that stars belonging to a given MG share similar spectroscopic properties 
 (e.g. Montes et al. 2001a; L\'opez-Santiago et al. 2009,2010).
 \nocite{2001A&A...379..976M,2009A&A...499..129L,2010arXiv1002.1663L}
 These studies exploit the many advantages of the nearby late-type stars. First, spectra of late-type stars
 are full of narrow absorption lines, allowing determination of accurate radial velocities.
 In addition, it is unlikely that an old star by chance shares chromospheric indices or
 a lithium abundance similar to those of  young solar-like stars,
 which provides means for assessing the likelihood of membership
 of a given star that are independent of its kinematics 
 \citep[e.g.][]{1993AJ....105..226S}.

 In this paper we present a search for classical MG members by analysing
 the kinematic and spectroscopic properties of a sample of nearby late-type stars.
 Section ~\ref{sample} describes the stellar sample and the observations and data reduction
 are described in Section ~\ref{observations}. A detailed analysis
 of the kinematic properties of the stars is given in Section ~\ref{kinematic}.
 Age indicators for solar-like stars are analysed in Section ~\ref{ages}. 
 A combination of the results from Sections ~\ref{kinematic} and ~\ref{ages} is used in Section
 ~\ref{finalmem} to analyse the MG membership of the stars.
 Section ~\ref{summary} summarizes our results.


\section{The stellar sample, observations, and data reduction}
\label{sample}

  Our  reference stellar  sample consists of  main-sequence (luminosity
  classes V/IV-V) FGK stars located at  distances less than 25 pc. The
  stars  have   been  selected   from  the  the   Hipparcos  catalogue
  \citep{1997ESASP1200.....P}, since it  constitutes a  homogeneous
  database  especially  for distance  estimates  - parallax
  errors are typically about 1 milliarcsec. In this work we have taken
  the   revised  parallaxes  computed by   \cite{Leeuwen}  from
  Hipparcos' raw data. No other selection criteria have been applied
  to the sample.\\

\begin{figure}
\centering
\includegraphics[angle=270,scale=0.5]{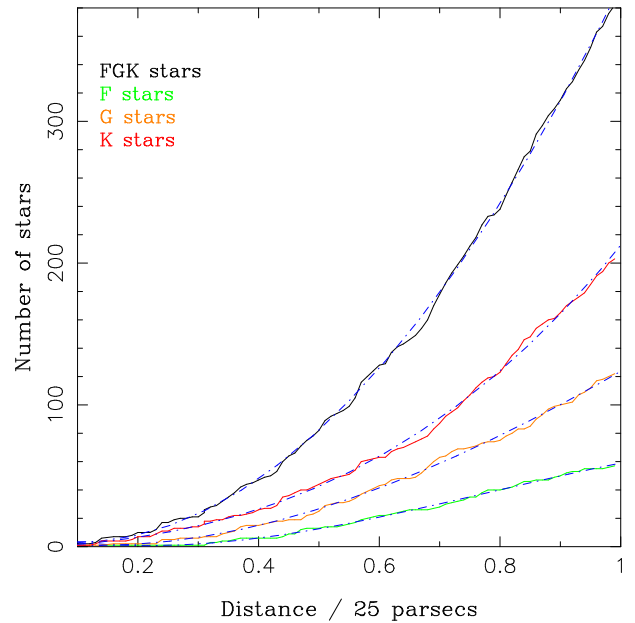}
\caption{Number of stars versus distance (normalized to 25 pc) for the F stars (green), G stars
(orange), K stars (red) and for the observed 405 stars. 
Fits to cubic laws are plotted in blue.}
\label{completitud}
\end{figure}

  The sample is most likely complete for FG-type stars; i.e., it 
  constitutes a volume-limited sample since the Hipparcos
  catalogue is complete for these spectral types. 
  In the case of K-type stars, Hipparcos is incomplete
  beyond  $\sim$ 15  pc; however, the
  number of the K-type stars is high enough for our purposes. 
  The final selection contains 126, 220, and 477 stars of spectral types F,G 
  and K respectively.
  In this contribution we present our first results for an
  observed subsample of 405 stars.
  The completeness of the observed sample can be seen in Figure ~\ref{completitud}
  where the number of objects is plotted as a function of distance, and the distribution
  fits well a cubic law, which indicates that they are homogeneously
  distributed.
  M-type stars have in principle been excluded from this study;
  nevertheless, six M-type stars, members or candidate member of
  MGs, which exhibit high levels of chromospheric activity
  and are suspected to be young, have been included in
  order to better understand the properties of such stellar groups.\\


\begin{figure}
\centering
\includegraphics[angle=270,scale=0.4]{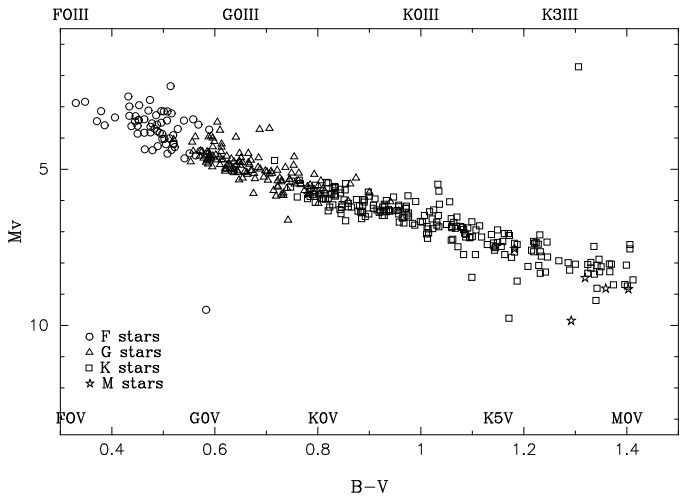}
\caption{HR Diagram for our sample of nearby late-type stars. F-type stars are plotted with circles; 
G-type stars with triangles; K-type stars with squares and M-type stars with stars.
}
\label{hrdiagram}
\end{figure}

  The observed stars are listed in Table~1, and 
  Figure ~\ref{hrdiagram} 
  shows the HR diagram of the sample.
  Several stars are clearly under the main sequence: HIP 4845, HIP 42525, HIP 49986, HIP 57939,
  HIP 72981, and HIP 96285.
  Hipparcos' spectral types for these stars are quite similar to those reported 
  in other catalogues such as \cite{Wright03}, \cite{Skiff09}, or SIMBAD. Only for HIP 72981 
  is incomplete, giving simply `K:', whereas SIMBAD gives M1 and
  the most updated reference in \cite{Skiff09} gives M2. However, the colour index $B-V = 1.17$ 
  suggests an early type, around K5. 
  HIP 42525 is a star in a double system and
  has a large uncertainty in the parallax ($\sigma_{\pi} = \pm 15.51 \ \rm{mas}$).
  Stars with uncertainties over 10 milliarcsec are identified with
  a symbol ${\dag}$ in Table~1. 
  The original selection (and therefore the observations) of the sample was made before the release of the 
  revised Hipparcos parallaxes \citep{Leeuwen}, and some of our stars are now out of
  the 25 pc distance because their revised parallaxes are slightly smaller. These
  stars are identified with a symbol ${\ddag}$ in Table~1. 
  The most ``extreme'' case is HIP 1692 whose parallax has changed from 43.42 $\pm$ 1.88 mas to
  3.23 $\pm$ 1.43 mas. This new parallax places the star in the giant branch as 
  is clearly shown in Figure ~\ref{hrdiagram} (square in the upper right corner).\\


\onllongtabL{1}{
\label{tabla1}
\begin{landscape}
\begin{scriptsize}
\begin{longtable}{llccccclcccccl}
\caption{Positions, proper motions and radial velocities for the observed stars.}\\
\hline
      &      &          &          &       &                                &                &                 &
\multicolumn{5}{c}{\textbf{Radial Velocity}}  & \\
      &      &          &          &       &                                &                &                 &
\multicolumn{5}{c}{\hrulefill}                &  \\
$HIP$ & $HD$ & $\alpha$ & $\delta$ & $\pi$ & $\mu_{\alpha}\times\cos\delta$ & $\mu_{\delta}$ & $Obs.run^{\sharp}$ &
$This work$ & $KH07$ & $NO04$ & $NI02$ & $VF05$  & Notes  \\
       &   & (h,m,s) & ($^{o}$,','')    & (mas) &  (mas/yr) & (mas/yr) & & (km/s) & (km/s) & (km/s) & (km/s) & (km/s) & \\
(1)    & (2) & (3) & (4) & (5) & (6) & (7) & (8) & (9) & (10) & (11) & (12) & (13) & (14) \\
\endfirsthead
\multicolumn{14}{c}
{{ \tablename\ \thetable{}  Continued}} \\
\hline
      &      &          &          &       &                                &                &                 &
\multicolumn{5}{c}{\textbf{Radial Velocity}}\\
      &      &          &          &       &                                &                &                 &
\multicolumn{5}{c}{\hrulefill}\\
$HIP$ & $HD$ & $\alpha$ & $\delta$ & $\pi$ & $\mu_{\alpha}\times\cos\delta$ & $\mu_{\delta}$ & $Obs.run^{\sharp}$ &
$This work$ & $KH07$ & $NO04$ &  $NI02$ & $VF05$ & Notes \\
       &   & (h,m,s) & ($^{o}$,','')    & (mas) &  (mas/yr) & (mas/yr) & & (km/s) & (km/s) & (km/s) & (km/s) & (km/s) & \\
(1)    & (2) & (3) & (4) & (5) & (6) & (7) & (8) & (9) & (10) & (11) & (12) & (13) & (14) \\
\hline
\endhead
\hline 
\endfoot
\hline \hline
\endlastfoot
\noalign{\smallskip}\hline\noalign{\smallskip}

171	&	224930	&	 00 02 10.1552	&	 +27 04 56.122  	&	82.18	$\pm$	2.23	&	829.90	$\pm$	1.20	&	-989.40	$\pm$	1.10	&	[9]	&	-37.90			&	-36.90	$\pm$	5.24	&	-36.90	$\pm$	0.40	&		&		&	Spec. Binary	\\
544	&	166	&	 	00 06 36.7839 	&	 +29 01 17.406  	&	73.16	$\pm$	0.56	&	380.50	$\pm$	0.90	&	-177.90	$\pm$	0.90	&	[8]	&	-6.66	$\pm$	0.07	&	-6.90	$\pm$	0.22	&	-6.90	$\pm$	0.10	&	-6.54	&	-5.8	&		\\
	&		&		&		&				&				&				&	[9]	&	-6.69	$\pm$	0.04	&				&				&		&		&		\\
910	&	693	&	 	00 11 15.8573 	&	 -15 28 04.719  	&	53.33	$\pm$	0.64	&	-82.50	$\pm$	0.90	&	-270.90	$\pm$	1.00	&	[8] 	&	16.70	$\pm$	0.08	&	14.90	$\pm$	0.71	&	14.90	$\pm$	0.10	&		&		&	 	\\
1499$^{\star}$	&	1461	&	 	00 18 41.8677 	&	 -08 03 10.805  	&	43.02	$\pm$	0.51	&	419.00	$\pm$	1.00	&	-145.10	$\pm$	0.90	&	[1]	&	-10.13	$\pm$	0.10	&	-10.49	$\pm$	0.43	&	-10.50	$\pm$	0.10	&	-10.17	&	-9.9	& 	\\
1532	&		&	 	00 19 05.5619 	&	 -09 57 53.474  	&	47.43	$\pm$	1.65	&	-36.50	$\pm$	1.60	&	-305.30	$\pm$	1.80	&	[4]	&	-10.88	$\pm$	0.23	&	-11.10	$\pm$	0.30	&				&		&		&		\\
1598	&	1562	&	 	00 20 00.4085 	&	 +38 13 38.629  	&	40.32	$\pm$	0.59	&	-141.80	$\pm$	1.00	&	-276.70	$\pm$	1.10	&	[1]	&	15.27	$\pm$	0.70	&	15.08	$\pm$	2.62	&	15.10	$\pm$	0.20	&		&		&		\\
1599	&	1581	&	 	00 20 04.2601 	&	 -64 52 29.246  	&	116.46	$\pm$	0.16	&	1708.40	$\pm$	0.80	&	1164.80	$\pm$	0.80	&	[10]	&	10.47	$\pm$	0.07	&	9.09	$\pm$	0.36	&	9.10	$\pm$	0.10	&		&	8.8	&		\\
\hline
\end{longtable}
\end{scriptsize}
\end{landscape}
}



\section{Observations and data reduction}
\label{observations}

 High-resolution spectra of 315 stars were obtained at the Calar Alto (Almer\'ia, Spain) 
 and La Palma (Canary Islands, Spain) observatories during eight observing runs. 
 Some stars (the most active ones)
 were observed  more than once. 
 The Calar Alto observations were taken with the fiber optics echelle spectrograph FOCES \citep{foces}
 attached at the Cassegrian focus of the 2.2 meter telescope. FOCES spectra have a resolution of $\sim\!57000$ 
 and cover a spectral range $\lambda \lambda$ 3800 - 10000 \AA. 
 La Palma observations were done at the 3.56 m Telescopio Nazionale Galileo (TNG) 
 using the cross dispersed echelle spectrograph SARG \citep{sarg}.
 In this case the resolution and  spectral range are $\sim\!57000$ and $\lambda \lambda$ 4960-10110 \AA, 
 respectively. Further details are given in Table ~\ref{log}. 

\begin{figure*}
 \centering
 \includegraphics[scale=0.15,angle=270]{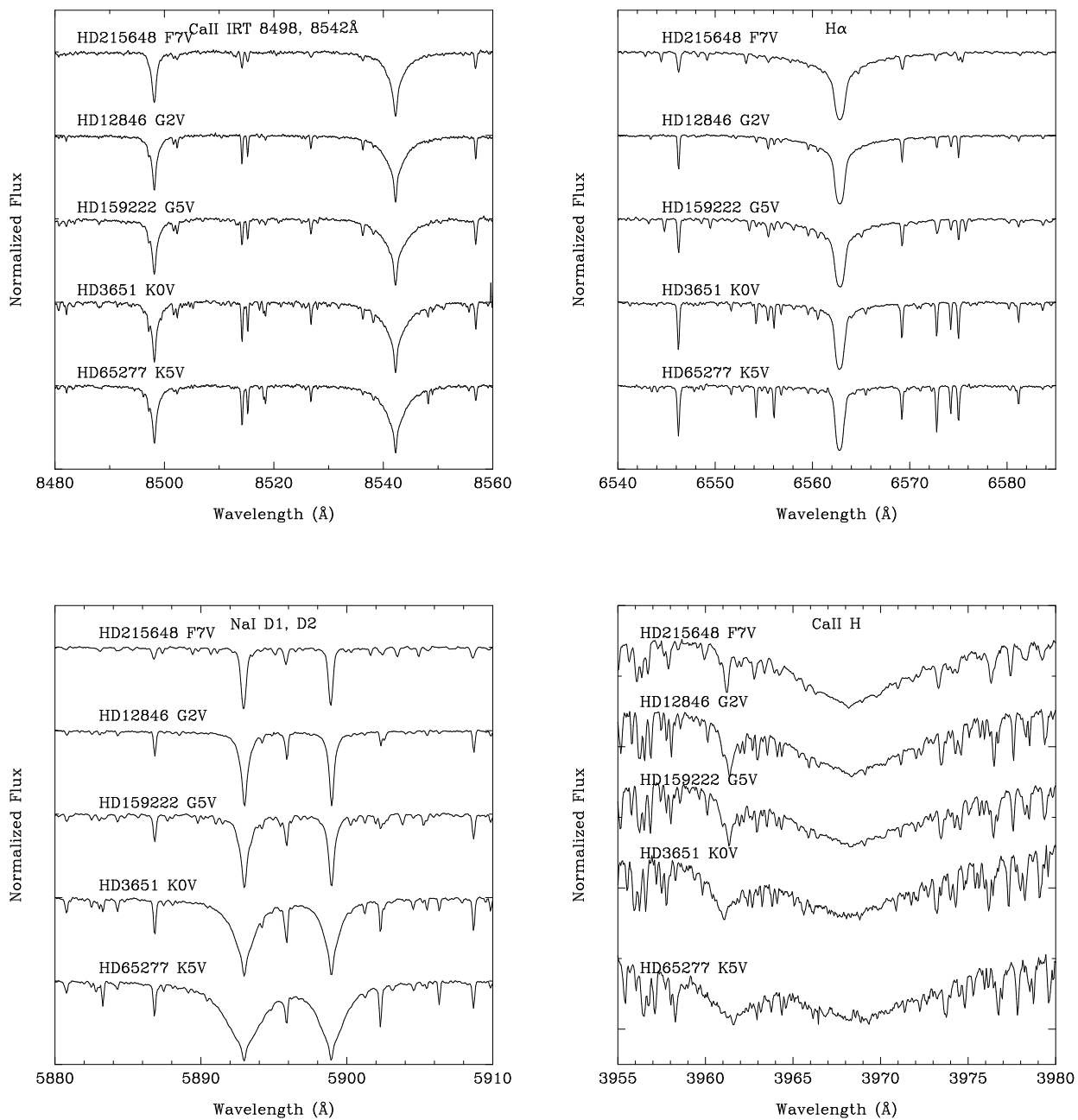}
 \caption{
 FOCES spectra of representative stars in the Ca~{\sc ii} IRT regions, $H\alpha$,
 Na~{\sc i} $D_{1}$, $D_{2}$, and Ca~{\sc ii} H \& K regions.}
 \label{spectra}
 \end{figure*}

 The spectra were reduced  using the standard procedures in the IRAF 
 \footnote{IRAF is distributed by the National Optical Astronomy Observatory, which is operated 
 by the Association of Universities for Research in Astronomy, Inc., under contract with the National Science Foundation.} 
 packages \textit{imred, ccdred,} and \textit{echelle}, i.e.  overscan, scattered light correction, and flat-fielding.
 Spectral orders were extracted  with the routine \textit{apall} and were normalized using 
 the IRAF task \textit{continuum} in order to compare the intensity of the lines and to measure
 equivalent widths. Thorium-Argon spectra were used for wavelength calibration.
 Figure ~\ref{spectra} shows examples of representative stars in different spectral regions.

 Since observations were done from northern observatories,
 most targets have $\delta > \rm{-25^{o}}$. Therefore additional spectra
 from public libraries have also been analysed. Specifically, 
 90  spectra were taken from the public library
 ``S$^{4}$N'' \citep{s4n}, which contains spectra taken  with the 2dcoud\'{e} spectrograph
 at Mc Donald Observatory and the FEROS instrument at the ESO 1.52 m telescope in La Silla. 
 Both the resolution and the spectral range are similar to  those of our own observations
 ($R\sim\!57000$, $\lambda \lambda$ 3620-9210 \AA). 
 FEROS spectra contribute partially to covering for the lack of southern targets.


\begin{table}
\centering
\begin{tiny}
\caption{Observing runs between 2005 and 2008.}
\label{log}
\begin{tabular}{cccccc}
\hline\noalign{\smallskip}
Date & Instrument & Spectral Range & orders & dispersion       & FWHM\\
     &            &  (\AA)         &        &  (\AA/pixel)     & (\AA)\\
\noalign{\smallskip}\hline\noalign{\smallskip}
 July 05 &  FOCES & 3470-10700  & 111 & 0.04-0.13 & 0.07-0.42  \\
 Jan 06  &  FOCES & 3470-10700  & 111 & 0.05-0.13 & 0.09-0.39  \\
 Feb 06  &  SARG  & 5600-10000  &  50 & 0.02-0.04 & 0.09-0.14  \\
 Dec 06  &  FOCES & 3640-10700  & 106 & 0.05-0.13 & 0.09-0.22  \\
 Jan 07  &  SARG  & 5600-10000  &  50 & 0.02-0.04 & 0.09-0.14  \\
 Apr 07  &  SARG  & 5600-10000  &  50 & 0.02-0.04 & 0.10-0.13  \\
 07A$^{\ddag}$    &  FOCES & 3470-10700    &  106  & 0.04-0.13  & 0.09-0.25 \\
 Nov 08  &  SARG  & 5600-10000  &  50 & 0.02-0.04 & 0.09-0.14  \\
\hline
\multicolumn{6}{l}{$\ddag$ First semester 07. Service Mode.}\\
\noalign{\smallskip}\hline\noalign{\smallskip}
\end{tabular}
\end{tiny}
\end{table}



\section{Kinematic analysis}
\label{kinematic}


\subsection{Radial velocities}
\label{radialv}

Radial velocities were measured by cross-correlating order by order,
using the IRAF routine \textit{fxcor}, the spectra of our programm stars with
spectra of radial velocity standard stars of similar spectral types (Table~\ref{tabla3}),
taken from \cite{1986PASP...98..223B}, \cite{1979PASP...91..698B}, and
\cite{1999ASPC..185..383U,1999ASPC..185..367U}.
Spectral orders with chromospheric features
and prominent telluric lines were excluded when determining the mean radial velocity.
Typical uncertainties are
between 0.15 and 0.25 km/s, while
maximum uncertainties are around 1-2 km/s.
Column~9 of Table~1, 
gives our results for the radial velocities.
A large number of stars in our sample (51) are known spectroscopic binaries
and are listed in  \textit{The 9th catalogue of spectroscopic binaries} 
\citep[][hereafter SB9]{SB9} and \textit{The 3rd Catalogue of Chromospherically 
Active Binary Stars} \citep{2008MNRAS.389.1722E}.
They are identified
in Table~1 
with the label `Spec. Binary'.
For those stars we have considered the radial velocity of the centre of mass of the system.\\

We have compared our results with radial velocity estimates by 
Kharchenko et al. (2007, hereafter KH07), Nordstr\"{o}m et al. (2004, hereafter NO04),
Valenti \& Fischer (2005, hereafter VF05), and Nidever  et al. (2002, hereafter NI02).
\nocite{2007AN....328..889K,2004A&A...418..989N,2005ApJS..159..141V,2002ApJS..141..503N}
These values are also given  
columns 10 to 13 of Table~1.
Of the 405 stars in our sample, 366 are found in  KH07, and
the differences among the radial velocity values in that work
and our results are less than 2 km/s for 290 stars, i.e., 79.2\% of the
common stars. 
A comparison with the NO04 data shows that, for 215 out of 
251 common stars (i.e. 85.6\%), the corresponding differences between the radial velocities
are less than 2 km/s.
A similar result, 85.3\% (177 out of the 151 common stars), is found
when considering VF05 data.
The comparison with NI02 is even better, because 179 out of 190 stars  (i.e. 94.2 \%) show
differences lower than 2 km/s.

\begin{figure}
\begin{minipage}{.48\linewidth}
\includegraphics[angle=270,scale=0.38]{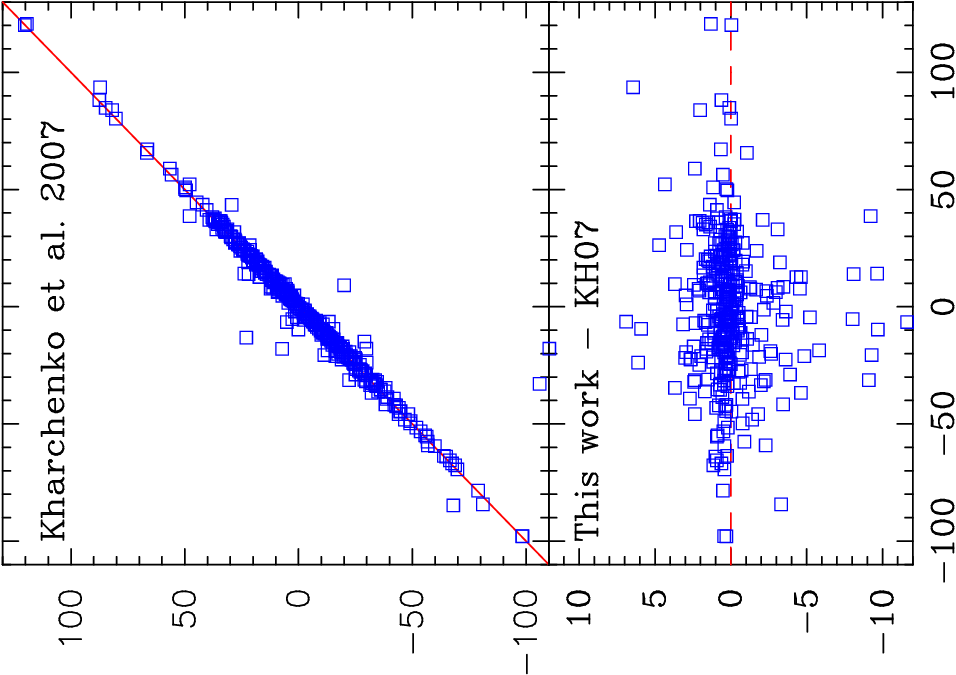}
\end{minipage}
\begin{minipage}{\linewidth}
\includegraphics[angle=270,scale=0.38]{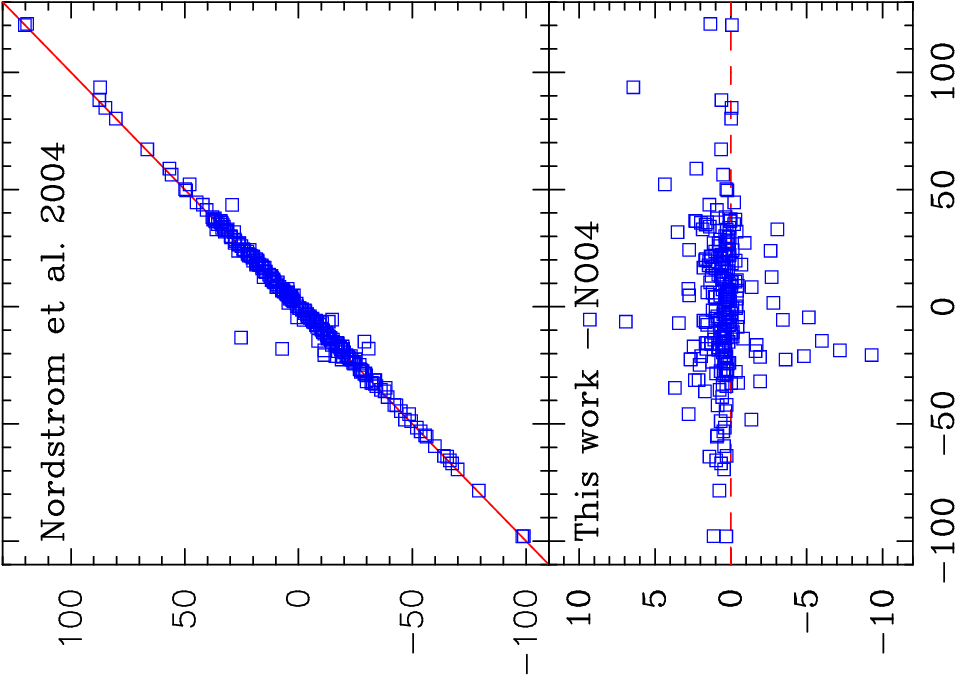}
\end{minipage}
\begin{minipage}{.48\linewidth}
\includegraphics[angle=270,scale=0.38]{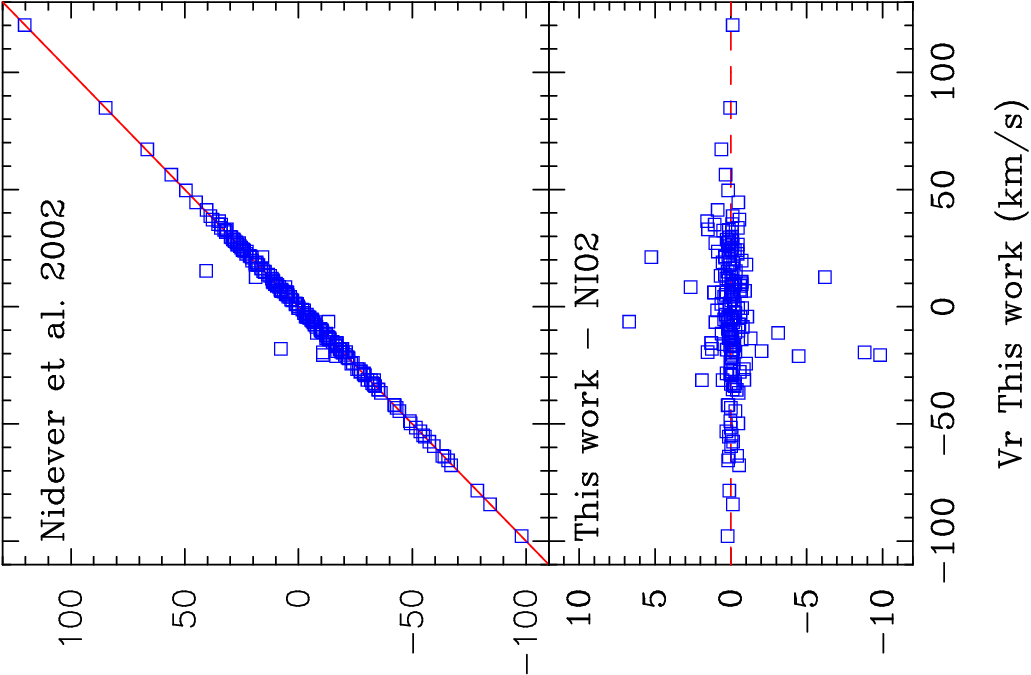}
\end{minipage}
\begin{minipage}{\linewidth}
\includegraphics[angle=270,scale=0.38]{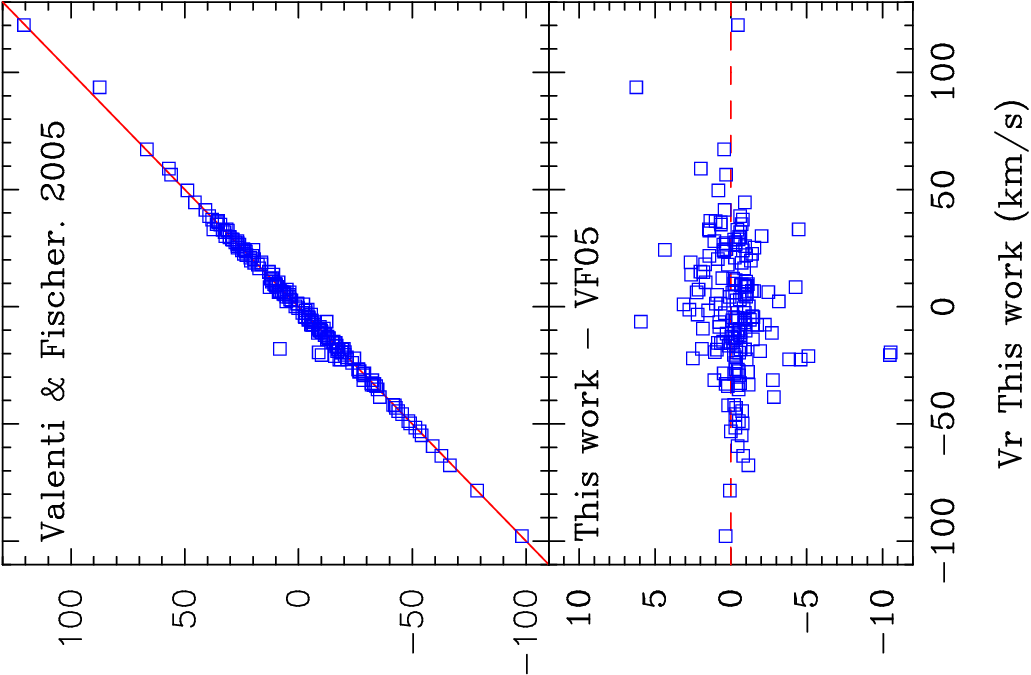}
\end{minipage}
\caption{
Comparison of radial velocities taken from the literature and obtained in this work.
Top left panel: \cite{2007AN....328..889K}; top right panel:\cite{2004A&A...418..989N};
bottom left panel:
\cite{2002ApJS..141..503N};
bottom right panel: \cite{2005ApJS..159..141V}
}
\label{figura4}
\end{figure}

Figure~\ref{figura4} illustrates these comparisons.
One can see that
the differences are
slightly greater with  KH07, likely because of the non-homogeneous origin of their radial
velocities values, mainly taken from \textit{The general catalogue or radial velocities} \citep{Barbier}.


\begin{table}
\caption{Radial velocity standard stars}
\label{tabla3}
\centering
\begin{tabular}{l c c c}
\hline\hline
Star & SpT & $V_{r}\pm\sigma_{V_{r}}$  & Reference \\
     &     &      (km/s)               &           \\
\hline
HD 102870 & F8V & $4.30$      & a \\
HD 50692 & G0V & $-15.05$     & a \\
HD 84737 & G0.5 & $6.0\pm1.1$ & b \\
HD 20630 & G5V & $18.0\pm1.0$ & b \\
HD 159222 & G5V & $-51.60$    & a \\
HD 82885 & G8III & $14.40$    & a \\
HD 65583  & G8V & $14.80$     & a \\
HD 144579 & G8V & $-59.45$    & a \\
HD 182488 & G8V & $-21.55$    & a \\
HD 102494 & G9IV & $-22.1\pm0.3$ & c \\
HD 62509 &  K0III & $3.2\pm0.3$  & c \\
HD 100696 & K0III & $0.2\pm0.5$  & b \\
HD 3651 &   K0V & $-32.96\pm0.8$ & b \\
HD 38230 &  K0V & $-29.25$      & a \\
HD 136442 & K0V & $-45.6\pm0.8$ & b \\
HD 92588 & K1IV & $43.5\pm0.3$  & d \\
HD 10476 &  K1V & $-33.9\pm0.9$ & b \\
HD 73667 & K1V & $-12.10$       & a \\
HD 124897 & K2III & $-5.3\pm0.3$ & c \\
HD 4628 &   K2V & $-10.1\pm0.4$  & d\\
HD 82106 & K3V & $29.75$   & a \\
HD 139323 & K3V & $-67.20$ & a  \\
HD 29139 & K5III & $54.29\pm0.2$ & c \\
\hline
a \cite{1999ASPC..185..367U}\\
b \cite{1986PASP...98..223B}\\
c \cite{1999ASPC..185..383U}\\
d \cite{1979PASP...91..698B}\\
\end{tabular}
\end{table}



\subsection{Identification of moving group candidates}
\label{mgcandidates}

 \cite{1993ApJ...402L...5S}  argued that, 
 in order to be convincingly classified as a kinematic group, a group of stars 
 should be moving through space at the same rate and in the same direction, and
 they should share the same velocity in the direction
 of the Galactic rotation $V$. This is because while motions in $U$ and $W$ lead to oscillations of
 the star about the mean motion of the group, diffusion in $V$ removes
 the star from its cohort forever.
 However, stars identified as group members show different structures tilted in the $(U,V)$ plane,
 i.e., do not form flat bars o ellipses with small $\sigma_{V}$ 
 \citep[e.g.][]{1997ESASP.402..525S},  and therefore both
 $U,V$ velocity components must be used to define more realistic
 membership criteria.\\

\begin{figure*}
\centering
\includegraphics[scale=0.45,angle=270]{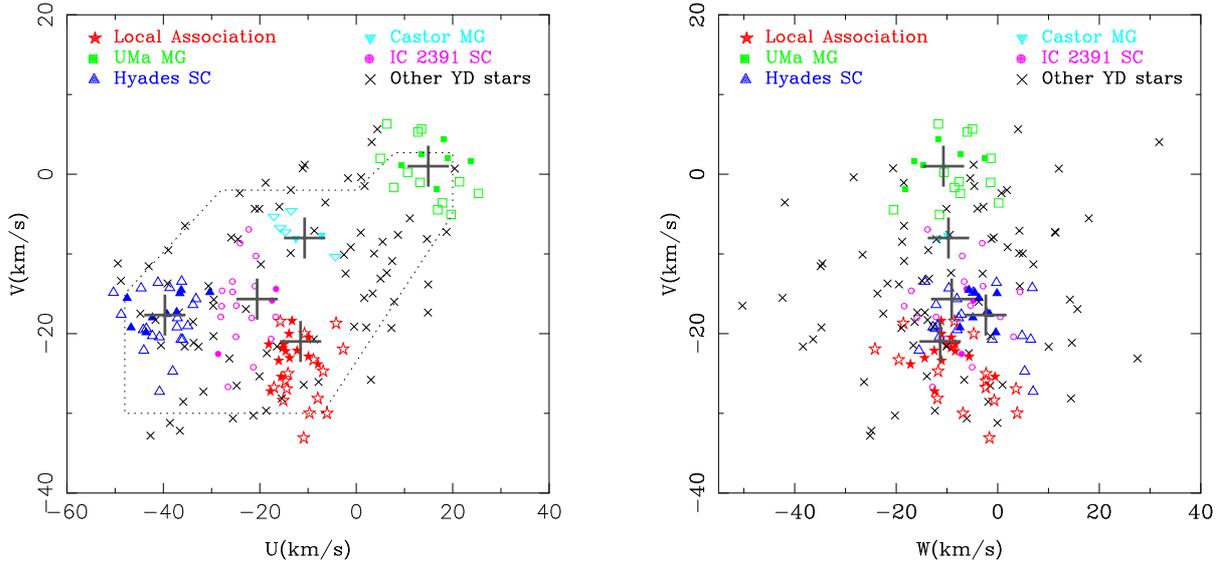}
\caption{ $(U,V)$ and $(W,V)$ planes for the observed stars.
Different colours and symbols indicate membership to different MGs.
Large crosses represent the convergence point of the young MGs shown in the figure.
The dotted line represents the boundary of the young disc population
as defined by Eggen \citep{1984AJ.....89.1358E,1989PASP..101..366E}.
Stars that satisfy both Eggen's criteria are shown with filled symbols, while
open symbols indicate stars that do not satisfy at least one of the Eggen's criteria.}
\label{uvwplane}
\end{figure*}

 Galactic spatial-velocity components $(U,V,W)$  were computed using our radial
 velocity results listed in Table~1, 
 together with Hipparcos parallaxes \citep{Leeuwen}
 and Tycho-2 proper motions \citep{2000A&A...355L..27H}.
 To compute $(U,V,W)$ we followed the procedure of \cite{2001MNRAS.328...45M}
 who updated the original algorithm of \cite{1987AJ.....93..864J} to epoch
 J2000 in the International Celestial Reference System (ICRS) as described
 in section 1.5 of The Hipparcos and Tycho Catalogues' \citep{1997ESASP1200.....P}. 
 To take the possible correlation between the astrometric
 parameters into account, the full covariance matrix was used in computing the
 uncertainties.
 To identify possible members of MGs we proceeded in two steps:

 \begin{itemize}

 \item \textit{i) Selection of young stars.} Young stars are assembled in a specific region
 of the $(U,V)$ plane with $(-50 \ \rm{km/s}\!<\!U\!<\!120 \ \rm{km/s};
 -30 \ \rm{km/s}\!<\!V\!<\!0 \ \rm{km/s})$, although the shape is not a square,
 see Figure~\ref{uvwplane}. 

 \item \textit{ii) Selection of possible members of MGs with small $V$ dispersion.}
 Considering previous results \citep{1997ESASP.402..525S,1999MNRAS.308..731S,2001MNRAS.328...45M},
 a dispersion of 8 km/s in the $U$, $V$ components  with respect to the central position
 of the MG in the $(U,V)$ plane is allowed. 
 The same dispersion is considered when taking the $W$ component into account. 

 \end{itemize}

 One hundred two stars of the sample have
 been classified as possible members of the different MGs:
 29 for the Local Association, 29 for the Hyades, 18 for the Ursa Major,
 19 for IC2391, and 7 for Castor.
 Column 2 of Table~\ref{eggen} lists these numbers, while the specific stars are
 listed in Tables~\ref{latable}  to~\ref{castable}.
 Their contents are described in Appendix~\ref{tables}.
 Another 78 stars have been selected as \textit{young disc stars}.
 These stars are inside or in the boundaries that determine the young disc population,
 but their
 possible inclusion in one of the stellar kinematic groups is not clear.
 The identified young disc stars are given in Table~14. 
 Figure ~\ref{uvwplane} shows the $(U,V)$ and $(W,V)$ planes, usually known as Bottlinger's diagrams, for these stars.\\


\subsection{Eggen's astrometric criteria}
\label{eggencriteria}

 To test whether a star ``belongs'' to a kinematic group,
 Eggen tried to establish ``strict'' criteria for MG-membership
 \citep{1958MNRAS.118...65E,1995AJ....110.2862E}.
 Eggen's criteria basically treat MGs, whose stars are extended in
 space, like open clusters whose stars are concentrated in space. Therefore, it is assumed
 that the total space velocities of the stars in the MG are parallel and move towards a common
 convergent point.
 The same relations of the moving-cluster method for the total and tangential
 velocities are applied, but taking into account only the components of
 the proper motion ($\mu$) oriented towards the convergence point ($\nu$) and the component
 of the proper motion
 oriented perpendicularly to the great circle between the star and the
 convergence point ($\tau$).
 The total ($V_{T}$) and tangential velocity (denoted as Peculiar Velocity, $PV$,
 by Eggen) can be combined to define a predicted radial velocity ($\rho_{c}$).\\
 
 The first membership criterion, namely the \textit{peculiar velocity criterion},
 is to compare the proper motion of the candidate to the proper motion expected
 if the star were a member of the MG; i.e., the candidate is accepted as an MG member
 if the ratio $\tau/\nu$ or $PV/V_{\rm Total}$ is ``sufficiently small''.
 \cite{1995AJ....110.2862E} considered a candidate to be a member if its
 peculiar velocity is less than 10\% of the total space velocity.\\

 The second membership criterion, the \textit{radial velocity criterion},
 compares the observed and the predicted radial velocities.
 \cite{1958MNRAS.118...65E} considered a star to be a member if the
 difference between both radial velocities is less than
 4-8 km/s.
 A more detailed discussion of these criteria can be found in
 \cite{2001MNRAS.328...45M}.\\

 Table ~\ref{eggen} gives the number of stars in each MG that satisfy
 both criteria (column 3), only the peculiar velocity criterion
 (column 4), and only the radial velocity criterion (column 5).
 Only a low percentage of the MGs members selected in the previous section
 satisfies both criteria (from $\approx$ 49\% in the Local Association
 to roughly 16\% in the IC2391 MG). 
 The results for individual stars are given in columns 8 and 9 of
 Tables ~\ref{latable} to ~\ref{castable}. For both $PV$ (column 8)
 and $\rho_{c}$ (column 9) criteria, there is a label, `Y' or `N', which indicates
 if the star satisfies the criteria.\\ 

 Eggen's criteria are not conclusive since they assumed a constant $V$ within the stars of
 a given MG. Anticipating some of the results in Section ~\ref{finalmem}, some stars for which
 both age estimates and $(U,V,W)$ components indicate that they are probable MG members do not
 satisfy these criteria.


\begin{table}
\centering
\begin{scriptsize}
\caption{Number of MGs candidates according to Eggen's criteria.}
\label{eggen}
\begin{tabular}{lcccc}
\hline\noalign{\smallskip}
Group &   Total stars & Both criteria & Only $PV$ & Only $\rho_{c}$ \\
\noalign{\smallskip}\hline\noalign{\smallskip}
Local Association  & 29 & 14 & 9  &  1  \\
IC2391             & 19 &  3 & 7  &  5  \\
Castor             &  7 &  2 & 0  &  4  \\
Ursa Major         & 18 &  6 & 1  &  7  \\
Hyades SC          & 29 &  9 & 5  &  9  \\
\noalign{\smallskip}\hline
\end{tabular}
\end{scriptsize}
\end{table}


\section{Age estimates}
\label{ages}

 Members of a given MG should be coeval and moderately young
 (only several Myr old, see Section ~\ref{introduction})
 therefore it is expected that MGs members share 
 age related-properties, such as similar chromospheric emission
 or lithium abundance. This provides the means of assessing the
 likelihood of membership for a given star that is independent
 of its kinematics.


\subsection{Lithium abundance}
\label{subsec51}

 Lithium abundance in late-type stars is a well-known age indicator
 since this element is destroyed as the convective motions gradually mix the stellar envelope
 with the hotter ($T\!\sim\!2.5 \times 10^{6}K$) inner regions.  
 However, it should only be regarded as an additional age indicator when
 compared with others since Li~{\sc i} equivalent width has a wide spread at a given
 age and mass, and consequently, the relation lithium-age is poorly constrained. Furthermore, for
 late K, M-type stars, lithium is burned so rapidly that it is only detectable for extremely young stars.
 Thus, the use of Li~{\sc i} as an age tracer is biased toward young stars, and
 it only provides low limits for  stars of the age of the Hyades or older.\\ 

 An  age estimate of the stars in our sample can be carried out by comparing their 
 Li~{\sc i} equivalent width,
 with those of stars in well known young open clusters of different ages
 \citep[e.g.][]{2001A&A...379..976M,Javi06}.
 Lithium \rm{EWs} have been obtained using the IRAF task \textit{sbands}, performing
 an integration within a band of 1.6 \space \AA \space centred in the lithium line.
 At the spectral resolution of our observations, the Li~{\sc i}  6707.8\AA \space line is blended
 with the Fe~{\sc i} 6707.41 \AA \space line.
 To correct for a possible contamination by Fe~{\sc i}, \cite{1990AJ.....99..595S}
 obtained an empirical relationship between
 the colour index $(B-V)$ and the  Fe~{\sc i} equivalent width, measured in stars that showed
 only the Fe~{\sc i} feature and no Li~{\sc i}.
 Soderblom's equation was obtained by using main-sequence and subgiant stars,
 so it does not account for possible luminosity-class effects. 
 Therefore we have built a new relationship, using only main-sequence stars without
 lithium detected in the spectrum:

 \begin{equation}
 \textrm{Fe~{\sc i} (EW)} = (0.020 \pm 0.005)(B-V) - (0.003\pm0.0015) (\AA).
 \end{equation}

 \noindent Which is fairly similar to the one obtained by Soderblom:

 \begin{equation}
 \textrm{Fe~{\sc i} (EW)} = 0.040(B-V) - 0.015 (\AA). 
 \end{equation}

 The EWs obtained are shown in column 10 in Tables ~\ref{latable} to ~\ref{castable},
 for each individual MG and in column 6 in Tables~14 to~15 
 for the stars classified as \textit{Other young disc stars}  and for the 
 stars not selected as possible MGs, respectively.

\begin{figure}
 \centering
 \includegraphics[scale=0.5,angle=270]{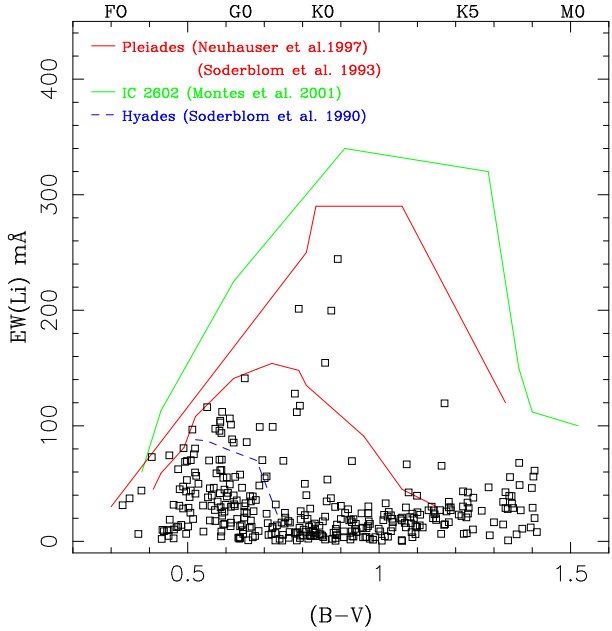}
 \caption{Li~{\sc i} vs $(B-V)$ diagram.
 Lines indicate the envelopes for the IC2602 (green), Pleiades (red), and Hyades (dashed blue).} 
 \label{lithium}
 \end{figure}

 Figure ~\ref{lithium} shows the \rm{EW} Li~{\sc i} versus colour index $(B-V)$ diagram.
 We have overplotted  the upper envelope of the Li~{\sc i} \rm{EW} of IC2602 (10-35 Myr)
 given by \cite{2001A&A...379..976M},
 the Pleiades cluster (78-125 Myr) upper envelope
 determined by \cite{1997A&A...325..647N}, and the lower envelope adopted by \cite{1993ApJ...402L...5S},
 as well as the Hyades open cluster (600 Myr) envelope adopted by \cite{1990AJ.....99..595S}.
 These clusters cover the range of ages of the MGs studied here (35-600 Myr).\\ 

 Nearly 4 \% of the stars are between the Pleiades envelopes, consistent with an age of
 $\sim$ 80 Myr. Roughly  8\% of the stars are between the Hyades and the
 Pleiades lower envelope with an age similar to those stars in the Ursa Major $\sim$ 300 Myr.
 Stars with lithium \rm{EW} below the Hyades envelope are likely to
 be older than 600 Myr. They are around the 23\% of the sample.
 Thus, approximately  35\% of the stars are moderately young (younger than 1 Gyr).
 Roughly 50\% of the stars lie below the  Pleaides lower envelope (but not below
 the Hyades' one). For these stars we can only state that they should be older than
 the Pleiades. Finally,
 stars with no photospheric Li~{\sc i} detected are expected to be older than 1 Gyr (around
 15\% of the whole sample).

 Concerning spectral types, the majority of the F stars are in the Hyades-like region
 of the diagram, with only five out of 61 
 stars in the Pleiades-like region.  For G-type stars, 71 out of 129
 are in the
 Hyades-like region, 34 in the Ursa Major-like region and only HIP 63742 shows an \rm{EW} comparable
 to those stars in the Pleiades. Finally,
 six out of 209 K-type stars are in the
 Pleiades-like region and 15 in the Ursa Major-like.\\ 
 
 The stars with the largest  Li~{\sc i} $EW$ are HIP 46816, HIP 46843, HIP 13402, HIP 63742
 HIP 75809, HIP 75829 (in this order). 
 According to their kinematics, HIP46816 has been classified in the \textit{young
 discs stars} category, whereas the three other stars have velocity-components
 $(U,V,W)$ in the boundaries of the Local Association
 (discussed in some detail in Section ~\ref{subsec61}).


\subsection{Stellar activity indicators}
\label{subsec52} 

 It is well known that for cool stars with convective outer-layers,
 chromospheric activity and  rotation are linked by the stellar dynamo
 \citep[e.g.][]{Kraft67,Noyes84,Benji} and both
 (activity and rotation)  diminish as the stars evolve. 
 Thus, activity/rotation tracers, such as $R'_{\rm HK}$, $L_{\rm X}$ or rotational
 periods are often used to estimate stellar ages
 \citep[for a recent detailed work on this subject see][]{2008ApJ...687.1264M}.

\subsubsection{Chromospheric emission: Ca~{\sc ii} H \& K lines}


 The stellar chromospheric activity is usually quantified by the $R'_{\rm HK}$ index,
 defined as the ratio of the chromospheric emission  in the
 cores of the broad Ca~{\sc ii} H \& K absorption lines to the total bolometric
 emission of the star \citep[e.g.][]{Noyes84}.
 The $R'_{\rm HK}$ values used in this work were taken from \cite{Raquel09}
 since they were obtained from the spectra in this paper.
 For those stars with no $R'_{\rm HK}$ value in \cite{Raquel09}, the $R'_{\rm HK}$ values have
 been taken from the literature (see references in Appendix ~\ref{tables}).\\

 Several relations between $\log R'_{\rm HK}$ and stellar chromospheric age are available in
 the literature \citep[e.g.][]{1991ApJ...375..722S}. In this paper we take those given by
 Mamajek \& Hillenbrand (2008, Eq.3):
 \nocite{2008ApJ...687.1264M}

 \begin{equation}
 \label{eqmamajek}
 \log(\tau/\rm{yr})=-38.053-17.912\log R'_{\rm HK}-1.6675 \log R_{\rm HK}^{'2}, 
 \end{equation}

 \noindent which is
 valid between $\log R'_{\rm HK}$ values of -4.0 and -5.1
 (i.e. $\log \tau$ of 6.7 and 9.9). Although the stars used in the calibration of
 Eq.~\ref{eqmamajek} are all stars with $(B-V) < 0.9$, we assume it holds for 
 the entire (B-V) range of our stars.   
 As in the case of the lithium abundance, activity indicators are also biased towards younger
 stars. The accuracy of Mamajek's relation is 15-20\% for young stars (younger than 0.5 Gyr),
 but beyond this age, uncertainties can grow up to more than 60\%. The
 $\log R'_{\rm HK}$ values and derived ages are shown in columns 11 and 12  in Tables ~\ref{latable}
 to ~\ref{castable}, and columns 7 and 8 in Tables ~14 to ~15.\\ 

\begin{figure}
\centering
\includegraphics[scale=0.5,angle=270]{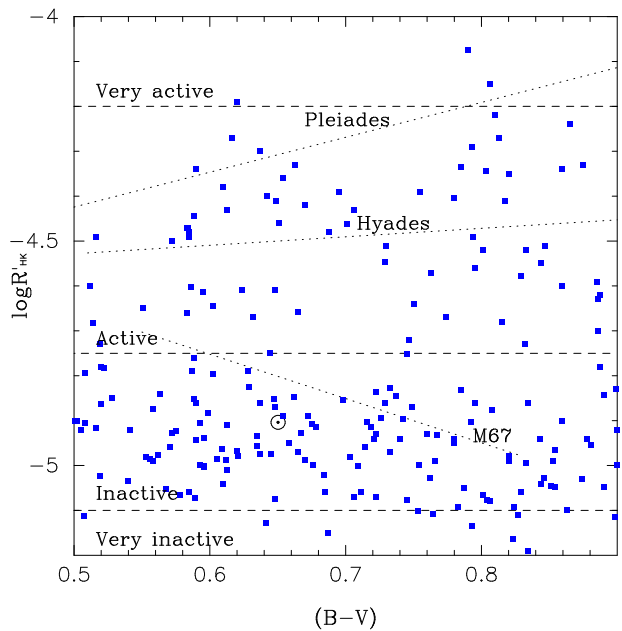}
\caption{ $\log R'_{\rm HK}$ vs $(B-V)$ colour.
The position of the Pleiades ($\sim$ 120 Myr), Hyades (600 Myr),
and M67 (4 Gyr) stars are indicated with dotted lines \citep{2008ApJ...687.1264M}.
The position of the Sun is also shown with a dotted circle. Dashed lines are the limits for
very active, active, inactive, and very inactive stars, according to \cite{1996AJ....111..439H}.
}
\label{calcio}
\end{figure}

 Figure ~\ref{calcio} shows the $\log R'_{\rm HK}$ versus $(B-V)$ diagram of stars in clusters of known ages.
 Following \cite{1996AJ....111..439H}, we used $\log R'_{HK}$ to classify
 stars into ``very inactive'' $(\log R'_{\rm HK}\!<\!-5.1)$, ``inactive'' $(-5.1\!<\! \log R'_{\rm HK} \!<\!-4.75)$,
 ``active''  $(-4.75\!<\! \log R'_{\rm HK}\!<\!-4.2)$, and ``very active'' if $\log R'_{\rm HK}\!>\!-4.2$.
 The percentages of stars in each region are 8\%, 47\%, 41\%, and 4\%, respectively.
 Mean $\log R'_{\rm HK}$ value for inactive stars is -4.93 with a standard deviation of 0.09,
 and $<\log R'_{\rm HK}> = -4.53$ with a standard deviation of 0.13
 for active stars. These numbers
 are quite similar to those found by \cite{1996AJ....111..439H} and \cite{Gray03}.\\

 Most of the stars in the ``very active'' category are, according to their
 kinematics, candidate members to MGs. HIP 46843 and HIP 86346 (Local Association),
 HIP 21482, and HIP 25220 (Hyades), HIP 8486 (Ursa Major), HIP 66252 (IC2391)
 HIP 33560 and HIP 46816 (young disc population).
 Three of the stars in this ``very active'' region, namely HIP 45963, HIP 21482 and
 HIP 91009 
 are well-known variable chromospherically active binaries
 (included in The 3rd Catalogue of chromospherically
 active binary stars \citep{2008MNRAS.389.1722E}).
 In those systems, stellar activity/rotation are enhanced by tidal interaction
 with the companion star, leading to high levels of chromospheric and coronal emission,
 up to two orders of magnitude higher than the level expected for a single star with
 the same rotation period \citep{1985ApJ...298..761B,1987ApJ...316..434S,1996A&A...312..221M}.
 Therefore their $\log R'_{\rm HK}$  values cannot provide any information on their
 age or membership to MGs. 
 Lithium abundance
 is also affected in this kind of systems, showing
 overabundances with respect to the typical values for single stars of the same mass and evolutionary
 stage \citep{1997A&A...326..780B}.


\subsubsection{Coronal emission: ROSAT data}

 In addition to their chromospheric activity, the rapid rotation
 of young stars drives a vigorous stellar dynamo, producing a strong, coronal
 X-ray emission. 
 Even though there are $L_{\rm X}$ values already published in several
 catalogues \citep[e.g.][]{Huensch}, in order to be self-consistent
 we have re-computed them with the revised Hipparcos parallaxes \citep{Leeuwen}
 used in this work.
 
 To compute $L_{\rm X}$,
 we searched for X-ray counterparts in the ROSAT All-Sky
 Survey Bright Source Catalogue \citep{1999A&A...349..389V} and the
 Faint Source Catalogue \citep{2000IAUC.7432R...1V}. To determine the X-ray fluxes
 we used the count rate-to-energy flux conversion factor ($C_{\rm X}$) relation given
 by \cite{1995ApJ...450..401F}:

 \begin{equation}
 C_{\rm X} = (8.31 + 5.30 \ \textrm{HR1})  10^{-12} \textrm{erg} \ \textrm{cm}^{-2} \ \textrm{counts}^{-1}.
 \end{equation}

 \noindent Where HR1 is the hardness ratio of the star in the ROSAT energy band 0.1-2.4 KeV.
 Combining the X-ray count rate, $f_{\rm X} (\textrm{counts \ s}^{-1})$, and the conversion factor $C_{\rm X}$ with
 the distance $D$, the stellar X-ray luminosity $L_{\rm X} (\textrm{erg\ s}^{-1})$ can be estimated:

 \begin{equation}
 L_{\rm X} = 4\pi D^{2} C_{\rm X} f_{\rm x}.
 \end{equation}

\begin{figure}
 \centering
 \includegraphics[scale=0.5,angle=270]{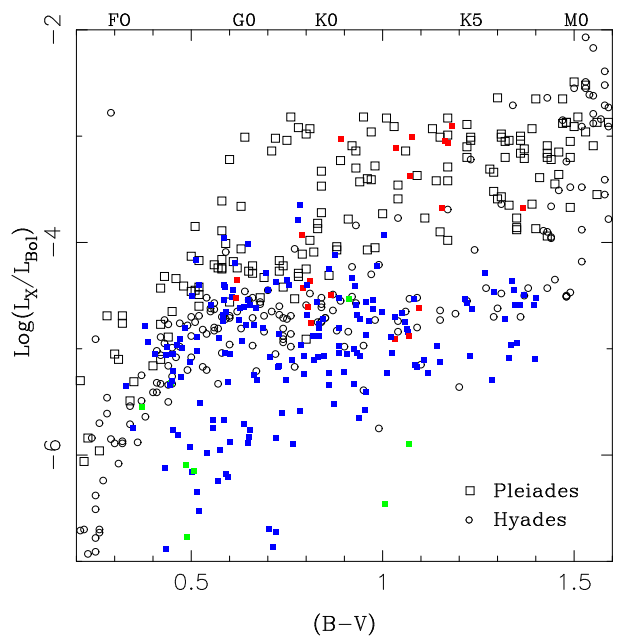}
 \caption{
 Fractional X-ray luminosity $\log (L_{\rm X}/L_{\rm Bol})$ vs colour index $B-V$.
 Stars classified as ``very active'' and ``very inactive'' according to their $\log R'_{\rm HK}$ 
 value are plotted in red and green colours, respectively.}
 \label{xray}
 \end{figure}

 Figure ~\ref{xray} shows the fractional X-ray luminosity 
 $L_{\rm X}/L_{ \rm Bol}$ versus the colour index $(B-V)$. 
 Bolometric corrections were derived from the $(B-V)$ colour by
 interpolating in Flower (1996, Table~3) \nocite{Flower96} 
 Data for the Pleaides \citep{1994ApJS...91..625S} and Hyades
 \citep{1995ApJ...448..683S} clusters have been overplotted
 for a comparison. Approximately  23\% of the stars are in the Pleiades
 region of the diagram, 51\% of the stars
 are in the Hyades region, and $\sim$ 26\% of the stars are below the Hyades' sequence.

 To compute the stellar age from the X-ray luminosity, we 
 followed the work by Garc\'es et al. (2010, in prep):
 
 \begin{equation}
 \label{lxage}
 \begin{array}{lr}
 L_{\rm X} = 6.3 \times 10^{-4} \ L_{\rm Bol}       &  (\tau<\tau_{i})\\
 L_{\rm X} = 1.89 \times 10^{28} \  \tau^{-1.55}   & (\tau>\tau_{i}) .
 \end{array}
 \end{equation}

 \noindent With $\tau_{i}=2\times 10^{20} L_{\rm Bol}^{-0.65}$, and
 both $L_{\rm X}$ and $L_{\rm Bol}$ are expressed in erg/s and $\tau$ is given in Gyr.
 Columns 13 and 14 in Tables ~\ref{latable}
 to ~\ref{castable} show the $L_{\rm X}/L_{\rm Bol}$ values and derived ages, while 
 in Tables ~14 to ~15 
 these data are in columns 9 and 10.\\

 The critical parameter $\tau_{i}$ marks the change from a non-saturated
 regime in which there is an inverse relation between the stellar rotation
 and $L_{\rm X}$ and the saturated regime in which the star reaches a maximum $L_{\rm X}$
 such that $L_{\rm X}/L_{\rm Bol} \approx 10^{-3} $ 
 \citep[e.g.][and references therein]{2003A&A...397..147P}.
 Only one star, HIP 86346, is in the saturated regime. For this star, 
 Eq.~\ref{lxage} only provides an upper limit to the age,
 close to the ``real'' age of the star. This star is discussed 
 in some detail in Section  ~\ref{subsec61}.\\

 Most of the stars included in the ``very inactive'' category defined before
 do not have ROSAT data, and for the few of them that do,
 X-ray data place them  below the Hyades' sequence (Fig~\ref{xray}).
 Lithium abundance shows a
 similar behaviour, and these stars are below the Hyades' envelope or do not show
 lithium at all. Although some of them have been identified by means of their kinematics
 as young disc stars or MG members, age
 diagnostics show that they are, however, old stars. 
 For the stars in the ``very active'' category, the situation is the opposite one. All of them
 have ROSAT data, and most of them have fractional X-ray luminosities similar to those of the Pleiades
 (Fig~\ref{xray}).
 They also show higher
 lithium abundances than the ``very inactive'' or the ``inactive'' stars.


 \subsubsection{Age from stellar rotation: Gyrochronology}

  Stars are born with relatively high rotational velocities.
  In the course of their evolution, rotation decreases due to the
  loss of angular momentum with stellar winds and magnetic breaking
  \citep{1967ApJ...148..217W,1993MNRAS.261..766J,2007A&A...473..501A}. Thus stellar rotation can
  be used to estimate stellar ages, and it is well known
  that solar-type stars follow a law of the form
  $P_{\rm Rot}  \propto t^{1/2}$ \citep{1972ApJ...171..565S}.
  Subsequent works have refined this relationship, e.g., by establishing a mass
  dependence in the evolution of  rotational periods
  \citep[e.g.][]{1989ApJ...343L..65K} or deriving a rotation-age relationship
  as a function of the stellar colour \citep{Barnes07,2008ApJ...687.1264M}.

  To compute ages, we follow the relationship
  given by \cite{2008ApJ...687.1264M}:

  \begin{equation}
  P_{\rm Rot}=0.407( (B-V) -  0.495)^{0.325} \times t^{0.566} .
  \end{equation}

  \noindent With the age of the star, $t$, given in Myr and the period in
  days.\\

\begin{figure}
 \centering
 \includegraphics[scale=0.5,angle=270]{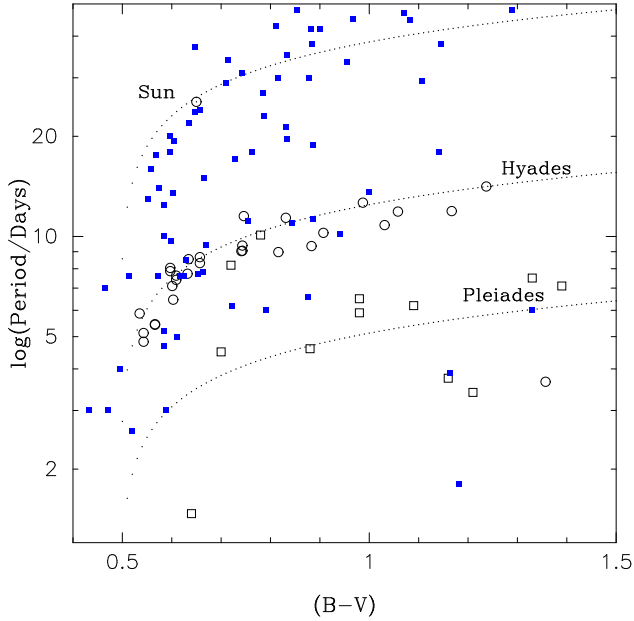}
 \caption{Rotation periods vs $(B-V)$ colour.
  Data from the Pleiades were taken from \cite{1995PASP..107..211P}, 
  whereas data from the Hyades are from \cite{1987ApJ...321..459R}.
  Three gyrochrones (at the ages of the Pleiades, Hyades, and the Sun) have been overplotted for
  a comparison.}
 \label{rotation}
 \end{figure}

  Rotational periods have been taken from 
  Noyes et al. (1984), Baliunas et al. (1996), Saar \& Osten (1997), and Messina
  et al. (2001).
  \nocite{Noyes84,Baliunas96,Saar97,Messina01}
  Unfortunately, only 17.3\% of the stars have measured rotational periods.
  Rotational periods and derived ages are
  given in columns 15 and 16  in Tables ~\ref{latable}
  to ~\ref{castable} and columns 11 and 12
  in Tables ~14 to ~ 15.  
  Figure~\ref{rotation} shows the rotation period
  for the stars of our sample as a function of the
  colour index $(B-V)$. Percentages of Pleiades-like, Hyades-like,
  and older stars
  are 22\%, 28\%, and 50\% respectively. 
  All stars with rotation periods lower than seven days are
  MGs candidates. The fastest rotator is HIP 86346
  with a period of only 1.8 days, while the slowest ones are HIP 3093
  and HIP 104217 with 48 days.


 \subsubsection{Discussion}

\begin{figure}
\includegraphics[scale=0.50,angle=270]{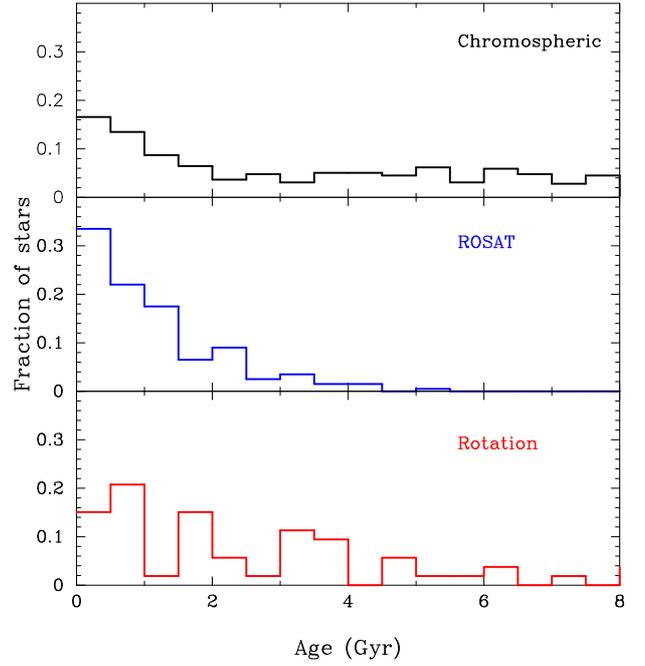}
\caption{
Age distribution for chromospheric-derived ages (black solid line), ROSAT ages
(blue line), and rotational ages (red line).}

\label{ageshistograms}
\end{figure}

  Figure ~\ref{ageshistograms} shows the age distribution for the
  different activity indicators.
  The results can be compared with those of  Mamajek \& Hillenbrand (2008, Figure 14).
  Chromospheric age  shows an enhancement of the star formation rate
  in the last 2 Gyr, then the distribution becomes more or less flat.
  We do not find a clear minimum at 2 Gyr, the so-called \textit{Vaughan-Preston gap}
  \citep{1980PASP...92..385V}. 
  ROSAT ages are biased towards stars younger than 3-4 Gyr;
  i.e., older stars have negligible (or undetectable) X-ray emission, and therefore
  their distribution does not offer information on the stellar formation history.
  As far as rotational ages are concerned, there are not enough stars with measured rotational
  periods to draw robust conclusions. 

  Although the agreement between the ROSAT and the chromospheric distribution
  is overall good, when considering individual stars there can be discrepancies,
  which can for to different reasons.
  For example, some stars present variability in their levels of activity,
  which leads to very different age estimates if the activity indicators
  are taken in different epochs of the activity cycle.
  For example, HIP 37349 is a known variable observed three times: $\log R'_{\rm HK}$ values
  are -4.54, -4.57, and -4.28 \citep{Raquel09}, which lead to ages 800, 950, and 115 Myr,
  respectively, while the ROSAT-derived age is 1.17 Gyr, which is compatible
  with 800-950 Myr but not with  115 Myr. 
  In addition, stellar rotation can be influenced
  by tidal interaction in binary systems, leading to completely
  different ages. Finally there could be other aspects like possible mismatches of X-ray sources
  with their optical counterparts.


\subsection{Additional criteria}
\label{subsec53}


\subsubsection{Presence of debris discs}
\label{secdebris} 
 
 It is now well established that debris discs are more common around young stars
 \citep[e.g.][]{Habing01,2004ApJ...613L..65Z,Siegler07}.
 As stars age they are on average orbited by increasingly fewer dust particles
 so a high value of the fractional dust
 luminosity, $f_{\rm d}$, can be used as an additional indicator of youth.  
 There is evidence that debris systems of high infrared luminosity
 are more intimately linked to young stellar kinematic groups than the
 majority of normal stars \citep[e.g][]{Moor06}; indeed, several of the
 stars with the strongest infrared excesses are members of MGs
 (e.g., $\beta$ Pic, Barrado y Navascu\'es et al. 1999).
 \nocite{1999ApJ...520L.123B}

 However, $f_{\rm d}$ is a rather inaccurate age diagnostic.
 First, the amount of excess emission shows large differences
 among stars within the same age range
 \citep[e.g][Fig.~7]{Siegler07}. Even though
 stars with significant excess emissions should in principle be young, 
 no further information can be given without additional age estimates.
 Moreover, there are relatively old systems (age $\gtrsim$  500 Myr) with high $f_{\rm d}$
 values ($f_{\rm d} \simeq 10^{-3} $) possibly associated with stochastic collisional
 events.

 Stars with known infrared excesses and
 their inclusion to MGs are given Table~\ref{debris}.  As shown in that table,
 the IR excesses of those stars are
 relatively moderate ($f_{\rm d}\!\simeq\!10^{-5}$) and most stars
 with excess are not related to MGs. For example,
 the three stars with
 the largest IR-excess (HIP 76375, HIP 40693, and HIP 32480) are old field
 stars. Both kinematics and activity-derived age confirm this.

\onltab{5}{
\begin{table}
\centering
\caption{Stars with known debris discs}
\label{debris}
\begin{tabular}{llcccc}
\hline\noalign{\smallskip}
HIP     &  HD     & $f_{d}$            &  Reference & MG  & Age     \\
        &         & $(10^{-5})$        &            &     & (Myr)   \\
\noalign{\smallskip}\hline\noalign{\smallskip}
544     &  166    &5.9          &[4]&   LA      &20-150  \\
1599    &  1581   &0.2-1.6      &[4]&           &        \\
13402   &  17925  &2.2-4.4      &[4]&   LA      &20-150  \\
15371   &  20807  &0.4-1.5      &[4]&           &        \\
16537   &  22049  &8.3          &[3]&           &        \\
16852   &  22484  &1.2-4.3      &[4]&   YD      &        \\
18859   &  25457  &10$\pm$2     &[2]&   LA      &20-150  \\
19335   &  25998  &2.7          &[1]&   HS      &600     \\
22263   &  30495  &2.0-3.0      &[4]&   IC2391  &35-55   \\
23693   &  33262  &0.2-1.1      &[4]&           &        \\
27072   &  38393  &0.77         &[3]&   UMa     &300     \\
27435   &  38858  &10           &[1]&   YD      &        \\
28103   &  40136  &2.04         &[3]&           &        \\
32480   &  48682  &11           &[1]&           &        \\
40693   &  69830  &20           &[2]&           &        \\
42430   &  73752  &3.21         &[3]&           &        \\
42438   &  72905  &0.6-1.5      &[4]&   UMa     &300     \\
43726   &  76151  &0.4-1.0      &[4]&   HS      &600     \\
51502   &  90089  &0.85         &[1]&           &        \\
62207   &  110897 &1.4-2.3      &[4]&           &        \\
64924   &  115617 &1.9-3.3      &[4]&           &        \\
65721   &  117176 &1.8-7.7      &[4]&           &        \\
71284   &  128167 &0.49         &[3]&           &        \\
71395   &  128311 &1.3-2.7      &[4]&   UMa     &300     \\
76375   &  139323 &78.6         &[3]&           &        \\
85235   &  158633 &4.1          &[1]&           &        \\
107350  &  206860 &0.6-1.5      &[4]&   LA      &20-150  \\
\hline
\multicolumn{6}{l}{[1] \space  \cite{Beichman06}; \space  [2] \space \cite{Moor06}}\\
\multicolumn{6}{l}{[3] \space  \cite{Rhee07};     \space  [4] \space \cite{Trilling08}}\\
\noalign{\smallskip}\hline\noalign{\smallskip}
\end{tabular}
\end{table}
}


 \subsubsection{Metallicity}

  Moving group members are supposed to have formed in the same
  molecular cloud, so, they should have similar metallicity. 
  Local Association and Ursa Major members are expected to have metallicities
  compatible with the solar value. 
  Recently, \cite{2009AJ....138.1292S} have obtained 
  $[\rm{Fe/H}] = +0.03$ for a sample of 20 Pleaides' stars with statistical
  and systematic uncertainties of $+0.002$ and $+0.05$, respectively.
  For members of the Ursa Major Group, 
  \cite{1990ApJ...351..467B} found
  $[\rm{Fe/H}] = -0.085, \ \sigma=0.021$. 
  Hyades' members should be slightly metal rich $[\rm{Fe/H}] = +0.14, \ \sigma=0.05$.
  Therefore we discard very metal-poor stars as ``good'' MG candidates.
  Considering that the ``old'' (2 Gyr)  and metal-rich 
  MG HR1614 has a mean metallicity of $[\rm{Fe/H}] = +0.19\pm0.06$ \citep{2000A&A...357..153F},
  stars with metal overabundances over $\approx +0.20$ should also be discarded.

  Reliable spectroscopic determinations of the metallicity for our stars were taken from
  the literature \citep{2008MNRAS.384..173F,2004AN....325....3F,2004A&A...415.1153S,
  2008A&A...487..373S, 2005PASJ...57..109T,2005ApJS..159..141V}. When no spectroscopic
  metallicities were found there, they were computed from Str\"{o}mgren
  indices \citep{Hauck97} by using the calibrations given by \cite{1989A&A...221...65S}.
  These values are given in columns 17 (Tables~\ref{latable}
  to~\ref{castable}) and 13 (Tables~14 to ~15). 

  An inspection of the metallicities obtained reveals that
  there are no MGs candidates among the most
  metal-poor stars. However, some MGs candidates
  have positive metallicities. This is especially evident in the Hyades MG where
  the 45\% of the candidates are more metal-rich than the Sun. Between them
  we find HIP 43587 and HIP 67275, which are among the most metal rich in our sample, 
  $\rm {[Fe/H] = +0.35}$, $\rm {[Fe/H] > +0.30}$, respectively, and they both are known
  to have planets. As we will see in Section ~\ref{subsec62}, their age estimates confirm that
  they are old stars and  not ``good'' Hyades' members.


\section{Comparison between kinematic and age estimates. Final membership.}
\label{finalmem}

 Tables~\ref{latable} to~\ref{castable} show a summary of the kinematic and
 spectroscopic properties, as well as age estimates, of the stars which are candidate
 members to the different MG, according to their $(U,V,W)$ velocity components.
 Each table refers to one specific MG. This summary classifies the
 MGs candidates into three different categories, which are similar to the ones by \cite{1993ApJ...402L...5S}:

 \begin{itemize}
 \item \textit{Probable non-member:}
 If the derived ages from the different indicators agree,  
 but they are in conflict with the object having an age as an MG member. 
 \item \textit{Doubtful member:}
 If there is important disagreement among the different age indicators,
 including here the assigned age of the corresponding MG,
 or there is lack of information (i.e., some age indicators are not available) 
 \item \textit{Probable member:}
 If age indicators agree and also do  with the position of the star in the $(U,V)$ plane. 
 \end{itemize}

 \noindent The following subsections describe the membership of the stars studied in this work and
 the properties of each MG individually


\subsection{Membership and properties of the Local Association candidates}
\label{subsec61}

 The concept of a Local Association of stars was introduced by Eggen
 \citep[e.g.][]{1975PASP...87...37E}. This association, also known as Pleiades
 MG or Pleiades stream, includes stars in the Pleiades, $\alpha$ Persei, and IC2602
 clusters, as well as stars in the Scorpius-Centaurus star-forming region. 
 In the past years, small associations or groups of very young stars have been detected
 among Local Association members (AB Dor, TW Hydrae,
 $\beta$ Pic, and others). The spatial motions of these new associations
 are quite similar, but they present a wide range in ages and distributions around
 the Sun \citep[e.g.][]{Zuckerman}, which leads to the question of whether it is
 reasonable to 
 consider the Local Association as a single entity.
 Addressing this problem is beyond the scope of this paper, so we consider
 the Local Association as a single MG.\\

 There are 29 stars (Table ~\ref{latable}) that have velocity components $(U,V,W)$ 
 consistent with the stars being  candidates to the Local Association.
 Eight out of the 29 stars do not satisfy other criteria; i.e., their age estimates
 suggest that they are older than 20-150 Myr commonly adopted for the Local Association,
 and we consider them to be non-members; Seven out of the 29 stars are considered as doubtful members
 while, 14 are good candidates, i.e., probable members.
 Although the number of candidates is too small to draw robust
 conclusions we can infer that the contamination by old main-sequence stars
 vary from roughly 25\% to  50\%.

 Table ~\ref{lacomparison} lists the candidates and our
 final classification (column 3) for the Local Association
 stellar membership. Some of our candidates have already
 been classified as members of the young association around the
 star AB Dor or members of the so-called Hercules-Lyra association 
 introduced by \cite{2004AN....325....3F}, which are listed in columns
 4 to 6 of Table ~\ref{lacomparison}.
 All the previously proposed members of AB Dor or Hercules-Lyra fall into
 our classification of probable Local Association
 members, with the only exception of  HIP 62523, which we have
 classified as a ``doubtful member''. Thirteen out of our 29 candidates
 have not been included in these previous studies. 

 Among the Local Association members there are some interesting stars:

  \begin{itemize}

  \item{\rm {HIP 13402 (HD 17925):}}
  Although it is classified as an RS CVn variable \citep{2008MNRAS.389.1722E},
  \cite{2001A&A...367..910C} shows that the binary hypothesis does not seem 
  to be consistent with the Hipparcos photometric data.
  The estimated EW Li~{\sc i} = 182.52 $\pm$ 4.63 m\AA  \space agrees with
  the 208 m\AA \space given by \cite{2001A&A...379..976M} and 
  the 197 m\AA \space given by \cite{1995A&A...295..147F}, which suggests 
  an age similar to the Pleaides ($\approx$ 80 Myr).
  In addition, the different age estimates agree very well and
  confirm that it is a young star; furthermore,  the star
  is known to have IR-excess at 70 $\mu$m, see Table ~\ref{debris} \citep{Trilling08}. 
  Thus, we consider that it is a
  reliable member of the Local Association.

  \item{\rm {HIP 18859 (HD 25457):}}
  This star is classified as a weak-line T Tauri  \citep[e.g.][]{2000A&A...356..157L}, and
  has a remarkable infrared excess of $f_{d}=1.0 \pm 0.2 \times 10^{-4}$, Table ~\ref{debris} \citep{Moor06}.
  The EW Li~{\sc i} and age estimates confirm its young evolutionary state.

  \item{\rm{HIP 86346 (HD 160934):}}
  This star is one of the few Hipparcos' M-type stars we have observed in this project.
  Available spectral types in the literature vary between K7 to M0 
  \citep{1995AJ....110.1838R,2004ApJ...613L..65Z}.
  This object is a flare star identified as a spectroscopic binary by \cite{2006Ap&SS.304...59G}.
  A close companion was detected using lucky imaging techniques by \cite{2007A&A...463..707H}.
  Our radial velocities vary between -25.37 and -28.39 km/s.
  In all epochs the main optical activity tracers (Ca~{\sc ii} IRT, $H_\alpha$, Na~{\sc i} $D_{1}$,
  $D_{2}$, Ca~{\sc ii} H \& K)  are in emission.
  This star is a very rapid rotator (for an M-type star) with $v\sin i$  between 21 and 23 km/s.
  Our \rm{EWs} Li~{\sc i} measurements vary from 11.1 to 55.7 m\AA \space and are agree
  with the 40 m\AA \space reported by \cite{2004ApJ...613L..65Z}.
  ROSAT-age indicates a star younger than 100 Myr
  (it is in the ``saturated'' regime in the $\log L_{\rm X}/L_{\rm  Bol}$ vs age diagram),
  the position of the star in a colour-magnitude diagram  $(M_{V} = 7.55 \pm 0.15;
  (V-I) = 2.58 \pm 0.91)$ suggests it is  a pre-main sequence star.

 \end{itemize}

\onltab{6}{
\begin{table}
\centering
\begin{small}
\caption{Comparison between our final membership for the
Local Association and previous studies.$^{\dag}$}
\label{lacomparison}
\begin{tabular}{llcccc}
\hline\noalign{\smallskip}
       &        &            &  \multicolumn{2}{c}{Hercules-Lyra} & AB Dor \\
HIP    & HD     & This work  &  FU04  & L06                       & ZU04   \\
\noalign{\smallskip}\hline\noalign{\smallskip}
544     &   166      &  Y   &   Y  &  Y  &     \\
3979    &   4915     &  N   &      &     &     \\
7576    &   10008    &  Y   &      &  Y  &     \\
7751    &   10360    &  N   &      &     &     \\
12929   &   17230    &  N   &      &     &     \\
13402   &   17925    &  Y   &   Y  &  ?  &     \\
18859   &   25457    &  Y   &      &     & Y   \\
19422   &   25665    &  ?   &      &     &     \\
26779   &   37394    &  N   &      &     &     \\
37288   &            &  ?   &      &     &     \\
46843   &   82443    &  Y   &      &  ?  &     \\
54155   &   96064    &  Y   &      &  ?  &     \\
54745   &   97334A   &  Y   &   Y  &     &     \\
57494   &   102392   &  ?   &      &     &     \\
62523   &   111395   &  ?   &   Y  &  ?  &     \\
63742   &   113449   &  Y   &   ?  &     & Y   \\
65515   &   116956   &  Y   &   Y  &   ? &     \\
69357   &   124106   &  N   &      &     &     \\
72146   &   130004   &  ?   &      &     &     \\
73695   &   133640   &  ?   &      &     &     \\
75809   &   139777   &  Y   &  Y   &   ? &     \\
75829   &   139813   &  Y   &  Y   &   ? &     \\
77408   &   141272   &  Y   &  Y   &   ? &     \\
79755   &   147379   &  N   &      &     &     \\
86346   &   160934   &  Y   &      &     & Y   \\
105038  &   202575   &  ?   &      &     &     \\
107350  &   206860   &  Y   &      &  Y  &     \\
108156  &   208313   &  N   &      &     &     \\
115341  &   220221   &  N   &      &     &     \\
\hline
\multicolumn{6}{l}{FU04: \cite{2004AN....325....3F};
ZU04:\cite{2004ApJ...613L..65Z}}\\
\multicolumn{6}{l}{L06: \cite{Javi06}}\\
\noalign{\smallskip}\hline\noalign{\smallskip}
\multicolumn{6}{l}{$^{\dag}$Label `Y' indicates
probable members, `?' doubtful members and}\\
\multicolumn{6}{l}{ `N' probable non-members,
respectively.}\\
\end{tabular}
\end{small}
\end{table}
}


\subsection{Membership and properties of the Hyades candidates}
\label{subsec62}

  The Hyades MG group or Hyades Supercluster
  \footnote{The terms moving group and supercluster are used here without
  distinction.}
  has a venerable history in the study of
  MGs since references to the Hyades MG group
  go back in time to the first works in this area
   \citep{1869RSPS...18..169P}.
  It is commonly related with the Hyades and Praesepe clusters, both of
  them with ages around 600 Myr. Recently, 
  \cite{2007A&A...461..957F} has
  found that the MG is in reality a mixture of two
  different populations: a group of coeval stars related
  to the Hyades cluster (the evaporating halo of the cluster)  and a second group of old stars
  with similar space motions.
  Age diagnostics analysed in Section ~\ref{ages} allow us, in principle,
  to distinguish between the two populations. 

  There are 29 stars  in the region of the $(U,V,W)$ planes 
  occupied by the Hyades MG. Eleven out of these 29 candidates have been classified
  as probable members \footnote{To avoid confusion we recall that by ``probable member''
  of the Hyades supercluster we mean member of the group of coeval stars evaporated from 
  the primordial Hyades cluster},
  nine as probable non-members, whereas the classification of the other
  nine stars remains unclear (Table ~\ref{hstable}).

 Our selection contains 14 stars in common with \cite{2010arXiv1002.1663L}.
 A comparison between our final classification and those 
 given by \cite{2010arXiv1002.1663L} is shown in Table ~\ref{hscomparison}.
 There is good agreement with two exceptions, 
 HIP 17420 (for which our age estimates suggest an old star)
 and  HIP 19335 (discussed below). 

 We briefly describe some interesting stars concerning this MG:

 \begin{itemize}

 \item{\rm{HIP 19335 (HD 25998):}}
  This F7V star has been identified as a T-Tauri star in the surroundings
  of the Taurus-Auriga star formation region
  \citep{1998A&AS..132..173L}, although it is located at a significantly shorter
  distance, 21 pc, than the commonly accepted distance of $\sim\!140$  pc to that star forming region.
  The Li~{\sc i} $EW$ = 93.1 $\pm$ 3.0 m\AA  \space confirms its youth 
  and agrees well with the rest of age indicators, between 96
  and 300 Myr. The star has infrared-excesses
  at both Spitzer 24 $\mu$m and 70 $\mu$m MIPS bands  \citep{Beichman06}.
  All this information also confirms the youth of this star,
  but it is likely too young to be a member of the Hyades MG.

 \item{\rm{HIP 43726 (HD 76151):}}
  The Li~{\sc i} \rm{EW} of 31.42 $\pm$ 3.66 m\AA \space of this star, as well as the estimated
  chromospheric and rotational ages of $\sim$ 1.0 Gyr, suggests that it is not a member of the Hyades MG.
  Interestingly, this is a relatively old star with a debris disc
  \citep{Trilling08,Beichman06}.

 \item{\rm{HIP 67275 (HD 120136, $\tau$ Boo):}}
  $\tau$ Boo is one of the first cases where an exoplanet was found \citep{1997ApJ...474L.115B}.
  There is a strong disagreement between the X-ray age estimate, 0.36 Gyr, and the
  chromospheric age, 4.78 Gyr. Li~{\sc i} $EW$ also suggests an old star.
  This agrees with other published ages, 1.3 Gyr \citep{2005ApJS..159..141V}, 2.1 Gyr
  \citep{2004A&A...418..989N}, and 2.52 Gyr \citep{2005A&A...443..609S}.
  It is therefore unlikely that HIP 67275 is a member of
  the Hyades MG.

 \end{itemize}

\onltab{7}{
\begin{table}
\centering
\caption{Comparison between our final memberships for the
Hyades MG and those given by \cite{2010arXiv1002.1663L}.$^{\dag}$}
\label{hscomparison}
\begin{tabular}{llcccc}
\hline\noalign{\smallskip}
HIP    & HD     & This work & LS10 \\
\noalign{\smallskip}\hline\noalign{\smallskip}
1803	&	1835	&	Y	&	Y	\\
4148	&	5133	&	N	&		\\
12709	&	16909	&	Y	&		\\
13976	&	18632	&	Y	&	?	\\
16134	&	21531	&	Y	&	?	\\
17420	&	23356	&	N	&	Y	\\
18774	&	24451	&	?	&		\\
19335	&	25998	&	N	&	Y	\\
21482	&	283750	&	?	&		\\
25220	&	35171	&	?	&	Y	\\
40035	&	68146	&	N	&		\\
42074	&	72760	&	Y	&	Y	\\
42333	&	73350	&	Y	&		\\
43587	&	75732	&	?	&	?	\\
43726	&	76151	&	N	&		\\
44248	&	76943	&	?	&		\\
46580	&	82106	&	Y	&		\\
47592	&	84117	&	N	&		\\
48411	&	85488	&	?	&	?	\\
63257	&	112575	&	?	&		\\
66147	&	117936	&	N	&		\\
67275	&	120136	&	N	&	?	\\
69526	&	124642	&	?	&		\\
72848	&	131511	&	Y	&		\\
90790	&	170657	&	N	&		\\
94346	&	180161	&	Y	&	?	\\
96085	&	183870	&	?	&	?	\\
104239	&	200968	&	Y	&	?	\\
116613	&	222143	&	Y	&	?	\\
\hline
\multicolumn{6}{l}{$^{\dag}$Label `Y' indicates
probable members, `?' doubtful}\\
\multicolumn{6}{l}{members and `N' probable non-members,
respectively.}\\
\end{tabular}
\end{table}
}


\subsection{Membership and properties of the Ursa Major moving group}
\label{subsec63}

 The concept of a group of stars sharing the same kinematic as Sirius goes back more
 than one century ago. Nowadays the group includes more than 100 stars 
 \citep{1992AJ....104.1493E,1993AJ....105..226S,
 2003AJ....125.1980K,2004AN....325....3F,Ammler}.
 Eighteen stars have velocity components $(U,V,W)$
 consistent with the star being a candidate for Ursa Major
 (Table~\ref{umatable}).
 Four out of the 18 stars do not satisfy other criteria, and their age estimates indicate
 that they are older than the 300 Myr commonly adopted for the Ursa Major MG.
 Another eight out of the 18 stars stars are considered as doubtful members,
 while six stars are probable members. 

 Table ~\ref{umacomparison} shows a comparison between
 our classification and those reported in the literature. There is good
 agreement specially in the stars classified as good members.
 Three candidates of this MG are of special interest:

 \begin{itemize}

  \item{\rm{HIP 42438 (HD 72905):}}
  This star is known to have infrared excesses at 60 and 70 $\mu$m
  \citep{Spangler01,Bryden06}.
  All age estimates agree with an age of $\approx 300$  Myr, which indicate
  that it is a probable member of the group.

  \item{\rm{HIP 71395 (HD 128311):}}
  This object is an example of a star with a planetary system 
  \citep{2003ApJ...582..455B,2005ApJ...632..638V}
  in a debris disc (excess at 70 $\mu$m found by Trilling 
  et al. 2008). \nocite{Trilling08}
  Our chromospheric-derived
  age of 430 Myr agrees with the 390 Myr given by  \cite{2005A&A...443..609S} and
  confirms that this star is a probable member of the Ursa Major group.

  \item{\rm{HIP 80337 (HD 147513):}}
  This star is also known to have a planet \citep{2004A&A...415..391M}. 
  Due to a problem with the  header's  spectra, no radial velocity could be
  obtained so we have adopted the value given by NO04.
  Measured Li~{\sc i} EW = 35.51 $\pm$ 3.5 m\AA \space suggests that it is older
  than the Hyades, which agrees with both chromospheric and rotational ages,
  around 700 Myr. However the ROSAT age is much shorter, only 370 Myr. Therefore, we have
  classified this star as a ``doubtful'' member. 

 \end{itemize}

\onltab{8}{
\begin{table}
\centering
\caption{Comparison between our final memberships for the
Ursa Major MG and those previously reported in the literature.
$^{\dag}$}
\label{umacomparison}
\begin{tabular}{llccccc}
\hline\noalign{\smallskip}
HIP    & HD     & This work & SO93  & KI03 & FU04& LS10 \\
\noalign{\smallskip}\hline\noalign{\smallskip}
5944   & 7590	& Y  &	    &	    &	Y  &   \\
8486   & 11131	& Y  &	Y   &	Y?  &	Y  & Y \\
27072  & 38393	& N  &	?   &	Y?  &	   & Y \\
27913  & 39587	& ?  &	Y   &	 Y  &	Y  &   \\
33277  & 50692	& N  &	N   &	?   &	   &   \\
36827  & 60491	& ?  &	    &	N?/?&	   & ? \\
37349  & 61606A	& ?  &	    &	N?  &	   & Y \\
42438  & 72905	& Y  &	Y   &	Y?  &	Y  & ? \\
60866  & 108581	& ?  &	    &	    &	   &   \\
71395  & 128311	& Y  &	    &	?   &	   & Y \\
72659  & 131156A& Y  &	Y   &	?   &	Y  & ? \\
73996  & 134083	& N  &	    &	N?  &	   & Y \\
74702  & 135599	& N  &	    &	?   &	Y  &   \\
80337  & 147513A& ?  &	Y   &  N?/? &	   & Y \\	
80686  & 147584	& ?  &	    &	Y   &	   &   \\
96183  & 184385	& ?  &	    &	    &	   &   \\
102485 & 197692	& Y  &	    &	    &	   &   \\
108028 & 208038	& ?  &	    &	    &	   &   \\
\hline
\multicolumn{7}{l}{S093: \cite{1993AJ....105..226S}; KI03: \cite{2003AJ....125.1980K}}\\
\multicolumn{7}{l}{FU04: \cite{2004AN....325....3F}; LS10: \cite{2010arXiv1002.1663L}}\\
\noalign{\smallskip}\hline\noalign{\smallskip}
\multicolumn{7}{l}{$^{\dag}$Label `Y' indicates
probable members, `?' doubtful members and}\\
\multicolumn{7}{l}{ `N' probable non-members,
respectively.}\\
\end{tabular}
\end{table}
}


\subsection{Membership and properties of the IC2391 moving group}
\label{subsec64}

 The identification of an MG related to the IC2391 cluster is from  
 \cite{1991AJ....102.2028E,1995AJ....110.2862E}. Most of the stars
 listed as members of this MG are in fact early-type star members of 
 the cluster. By using the member's position in colour-magnitude diagrams
 Eggen obtained an age of $\sim$ 100 Myr, within an interval 
 spreading from 80 to 250 Myr.
 Recently, \cite{2010arXiv1002.1663L} has suggested the presence
 of two subgroups mixed in the $(U,V)$ plane with ages of 200-300
 and 700 Myr.\\

 Table~\ref{ictable} summarizes our membership criteria for the IC2391 MG. 
 Five out of 19 candidate stars have been classified as probable members, 10 as
 doubtful, and four as probable non-members.
 Our sample contains three stars in common with \cite{2010arXiv1002.1663L}. We confirm that 
 HIP 11072 is a doubtful member and HIP 59280 is a member of the old subgroup,
 but our age estimates for HIP 25119 disagree with \cite{2010arXiv1002.1663L}. We therefore
 consider this star as a ``non'' member instead of a member of the old
 subgroup since both chromospheric and ROSAT ages are around 3 - 5 Gyr.
 
 We have identified  four new stars as probable-members of the young subgroup: 
 HIP 19076, HIP 22263, HIP 29568, and HIP 66252. 
 In addition, HIP 71743 has been classified as a probable member
 of the old subgroup.  
 HIP 66252 (HD 118100, EQ Vir) is known to have flares, and therefore
 chromospheric and ROSAT ages can be greater than our estimates, although
 the lithium abundance confirms that it is a young star.
 HIP 34567 (ages between
 320 and 470 Myr)  should remain in the ``doubtful category''
 since it is a known chromospherically
 active binary \citep{2008MNRAS.389.1722E}.


\subsection{Membership and properties of the Castor moving group}
\label{subsec65}

 The Castor MG was originally suggested by \cite{1991A&A...252..123A}.
 This group includes, among other stars, three spectroscopic binaries (Castor A,
 Castor B, and YY Gem) and two prototypes of the $\beta$ Pic stars (Vega and
 Fomalhaut). \space \cite{Barrado} estimated an evolutionary age for this association 
 of 200 $\pm$ 100 Myr.\\

\begin{figure}
\centering
\includegraphics[angle=270,scale=0.5]{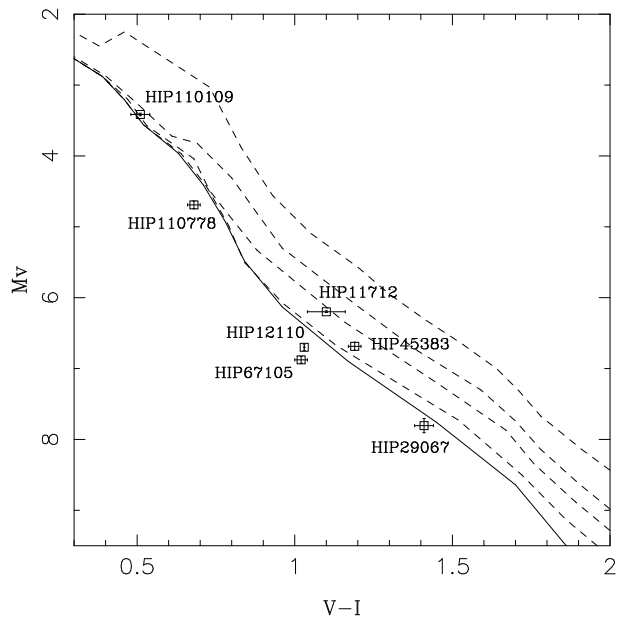}
\caption{
Colour-magnitude diagram for the Castor MG candidates.
Pre-main sequence isochrones from \cite{2000A&A...358..593S}
are plotted at 10, 20, 30 and 50 Myr.}
\label{cmd}
\end{figure}

 Only seven stars have been identified on the basis of their kinematics as candidate
 members of this group (Table ~\ref{castable}).
 Four were classified as probable members and three as doubtful members.
 HIP 29067 and HIP 109176 have been previously studied in detail by \cite{Barrado},
 and since we obtain similar results, we concentrate on the rest of candidates: 
 
 \begin{itemize}
 
 \item{\rm{HIP 12110 (HD 16270):}}
  There is a strong discrepancy between ROSAT and chromospheric ages, therefore
  the star remains as a doubtful member.

 \item{\rm{HIP 45383 (HD 79555):}}
  This star is a long-period astrometric binary \citep{WDSC}. Both ROSAT and chromospheric
  ages agree with the star being coeval with Castor MG members. As an additional
  test of youth, we plotted the star in a $M_{V}$ vs  $(V-I)$
  diagram (Figure ~\ref{cmd}).
  The  position of the star in this diagram suggests an age around 35 Myr. 
  Therefore we conclude that HIP 45383 is a young star and a probable Castor member. 
 
 \item{\rm{HIP 67105 (HD 119802):}}
  There is a strong discrepancy between ROSAT and chromospheric ages, therefore
  the star remains as doubtful member.

 \item{\rm{HIP 110778 (HD 212697):}}
  Both ROSAT and calcium  ages agree with the star being coeval with the Castor MG.
  Since this is a star in a multiple system, we have confirmed its youth
  nature by using colour-magnitude diagrams.  

 \item{\rm{HIP 117712 (HD 22378):}}
  This star is a known spectroscopic binary.
  Chromospheric ages suggest a moderately young star, between 600 and 860
  Myr. The position of the star in colour magnitude diagrams suggests
  also it is a young star, and hence a probable Castor MG member.

 \end{itemize}


\onltab{9}{
\begin{landscape}
\begin{table}
\centering
\begin{tiny}
\caption{Membership criteria for the Local Association candidate stars. 
(Convergence Point: 5.98 hours, -35.15 degrees; $U$ =-11.6 km/s, $V$ = -21.0 km/s, $W$ = -11.4 km/s;
Age: 20-150 Myr)
}
\label{latable}
\begin{tabular}{lcccccccccccccccc}
\noalign{\smallskip}\hline\noalign{\smallskip}
\hline\noalign{\smallskip}
       &  & \multicolumn{7}{c}{\textbf{Kinematics}} &    \multicolumn{1}{c}{\textbf{Lithium}}  & \multicolumn{2}{c}{\textbf{CaII H,K}} & \multicolumn{2}{c}{\textbf{ROSAT-data}} & \multicolumn{2}{c}{\textbf{Rotation}}  &   \\
       & &  \multicolumn{7}{c}{\hrulefill} &   \multicolumn{1}{c}{\hrulefill} & \multicolumn{2}{c}{\hrulefill} & \multicolumn{2}{c}{\hrulefill} &   \multicolumn{2}{c}{\hrulefill}  &   \\
$HIP$ & $(B-V)$ &  $U$    & $V$    & $W$    &  $V_{Total}$ & $V_{T}$ & $PV^{\star}$ & $\rho_{c}^{\star}$ &  EW(LiI)  &  $R'_{HK}$ & Age & $ L_{X}/L_{Bol}$ &   Age     & Prot     &  Age & 
[Fe/H]\\  
    &  & (km/s)  & (km/s) & (km/s) & (km/s)      & (km/s)   & (km/s)  & (km/s)  &   (m\AA)  &  (log) & (Gyr) & (log)          &    (Gyr)  &  (days) & (Gyr) &       \\
(1) & (2) & (3) & (4) & (5) & (6) & (7) & (8) & (9) & (10) & (11) & (12) & (13) & (14) & (15) & (16) & (17)\\

\noalign{\smallskip}\hline\noalign{\smallskip}
\multicolumn{17}{c}{\textit{Probable Non-members}}\\
\noalign{\smallskip}\hline\noalign{\smallskip}
3979	&	0.68	&	-15.58	$\pm$	0.32	&	-25.43	$\pm$	0.42	&	-0.53	$\pm$	0.18	&	29.83	&	29.83	&	5.48	Y	&	-7.01	Y	&	35.84	$\pm$	3.72	&	-4.92	[4]	&	5.04	&		&		&			&		&	-0.19	\\
7751	&	0.88	&	-4.19	$\pm$	0.16	&	-18.73	$\pm$	0.14	&	-18.60	$\pm$	0.07	&	26.72	&	15.16	&	-0.06	Y	&	10.19	N	&	2.14	$\pm$	4.65	&	-4.94	[1]	&	5.49	&		&		&	30	[13]	&	3.46	&	-0.23	\\
\end{tabular}
\end{tiny}
\end{table}
\end{landscape}
}
%
%
%

\onltab{10}{
\begin{landscape}
\begin{table}
\centering
\begin{tiny}
\caption{Membership criteria for the Hyades candidate stars.
(Convergence Point: 6.40 hours, 6.50 degrees; $U$ =-39.7 km/s, $V$ = -17.7 km/s, $W$ = -2.4 km/s;
Age: 600 Myr)
}
\label{hstable}
\begin{tabular}{lcccccccccccccccc}
\noalign{\smallskip}\hline\noalign{\smallskip}
\hline\noalign{\smallskip}
       &  & \multicolumn{7}{c}{\textbf{Kinematics}} &    \multicolumn{1}{c}{\textbf{Lithium}}  & \multicolumn{2}{c}{\textbf{CaII H,K}} & \multicolumn{2}{c}{\textbf{ROSAT-data}} & \multicolumn{2}{c}{\textbf{Rotation}}  &   \\
       &  & \multicolumn{7}{c}{\hrulefill} &   \multicolumn{1}{c}{\hrulefill} & \multicolumn{2}{c}{\hrulefill} & \multicolumn{2}{c}{\hrulefill} &   \multicolumn{2}{c}{\hrulefill}  &   \\
$HIP$ & $(B-V)$ &  $U$    & $V$    & $W$    &  $V_{Total}$ & $V_{T}$ & $PV^{\star}$ & $\rho_{c}^{\star}$ &  EW(LiI)  &  $R'_{HK}$ & Age & $ L_{X}/L_{Bol}$ &   Age     & Prot     &  Age & 
[Fe/H]\\  
    &  & (km/s)  & (km/s) & (km/s) & (km/s)      & (km/s)   & (km/s)  & (km/s)  &   (m\AA)  &  (log) & (Gyr) & (log)          &    (Gyr)  &  (days) & (Gyr) &     \\
(1) & (2) & (3) & (4) & (5) & (6) & (7) & (8) & (9) & (10) & (11) & (12) & (13) & (14) & (15) & (16) & (17)\\

\noalign{\smallskip}\hline\noalign{\smallskip}
\multicolumn{17}{c}{\textit{Probable  Non-members}}\\
\noalign{\smallskip}\hline\noalign{\smallskip}
4148	&	0.90	&	-36.34	$\pm$	0.32	&	-20.72	$\pm$	0.20	&	6.50	$\pm$	0.06	&	42.33	&	41.64	&	-4.63	N	&	-2.13	Y	&				&	-4.83	[1]	&	3.64	&	-5.22	&	2.04	&			&		&	-0.15	\\
17420	&	0.93	&	-30.35	$\pm$	0.18	&	-14.81	$\pm$	0.12	&	-4.58	$\pm$	0.16	&	34.08	&	31.28	&	-0.89	Y	&	21.33	Y	&	4.20	$\pm$	4.91	&	-5.04	[1]	&	7.35	&	-5.04	&	1.45	&			&		&	-0.12	\\
\end{tabular}
\end{tiny}
\end{table}
\end{landscape}
}
%
%
%
\onltab{11}{
\begin{landscape}
\begin{table}
\centering
\begin{tiny}
\caption{Membership criteria for the Ursa Major MG candidate stars.
(Convergence Point: 20.55 hours, -38.10 degrees; $U$ =14.9 km/s, $V$ = 1.0 km/s, $W$ = -10.7 km/s;
Age: 300 Myr)
}
\label{umatable}
\begin{tabular}{lcccccccccccccccc}
\noalign{\smallskip}\hline\noalign{\smallskip}
\hline\noalign{\smallskip}
       &  & \multicolumn{7}{c}{\textbf{Kinematics}} &    \multicolumn{1}{c}{\textbf{Lithium}}  & \multicolumn{2}{c}{\textbf{CaII H,K}} & \multicolumn{2}{c}{\textbf{ROSAT-data}} & \multicolumn{2}{c}{\textbf{Rotation}}  &   \\
       &  & \multicolumn{7}{c}{\hrulefill} &   \multicolumn{1}{c}{\hrulefill} & \multicolumn{2}{c}{\hrulefill} & \multicolumn{2}{c}{\hrulefill} &   \multicolumn{2}{c}{\hrulefill}  &   \\
$HIP$ & $(B-V)$ &  $U$    & $V$    & $W$    &  $V_{Total}$ & $V_{T}$ & $PV^{\star}$ & $\rho_{c}^{\star}$ &  EW(LiI)  &  $R'_{HK}$ & Age & $ L_{X}/L_{Bol}$ &   Age     & Prot     &  Age & 
[Fe/H]\\  
    &  & (km/s)  & (km/s) & (km/s) & (km/s)      & (km/s)   & (km/s)  & (km/s)  &   (m\AA)  &  (log) & (Gyr) & (log)          &    (Gyr)  &  (days) & (Gyr) &    \\
(1) & (2) & (3) & (4) & (5) & (6) & (7) & (8) & (9) & (10) & (11) & (12) & (13) & (14) & (15) & (16) & (17) \\
\noalign{\smallskip}\hline\noalign{\smallskip}
\multicolumn{17}{c}{\textit{Probable  Non-members}}\\
\noalign{\smallskip}\hline\noalign{\smallskip}
27072	&	0.49	&	18.18	$\pm$	0.13	&	4.44	$\pm$	0.13	&	-11.67	$\pm$	0.09	&	22.06	&	20.82	&	1.62	Y	&	-6.32	Y	&	68.82	$\pm$	2.90	&	-4.78	[7]	&	2.93	&		&		&			&		&	-0.09	\\
33277	&	0.60	&	13.53			&	5.66			&	-4.98			&	15.48	&	2.72	&	3.47	N	&	-2.48	N	&	15.28	$\pm$	3.35	&	-4.94	[3]	&	5.45	&		&		&			&		&	-0.13	\\
\end{tabular}
\end{tiny}
\end{table}
\end{landscape}
}
%
%
%
%
\onltab{12}{
\begin{landscape}
\begin{table}
\centering
\begin{tiny}
\caption{Membership criteria for the IC2391 MG candidate stars.
(Convergence Point: 5.82 hours, -12.44 degrees; $U$ =-20.6 km/s, $V$ = -15.7 km/s, $W$ = -9.1 km/s;
Age: 35-55 Myr)
}
\label{ictable}
\begin{tabular}{lcccccccccccccccc}
\noalign{\smallskip}\hline\noalign{\smallskip}
\hline\noalign{\smallskip}
    &   &  \multicolumn{7}{c}{\textbf{Kinematics}} &    \multicolumn{1}{c}{\textbf{Lithium}}  & \multicolumn{2}{c}{\textbf{CaII H,K}} & \multicolumn{2}{c}{\textbf{ROSAT-data}} & \multicolumn{2}{c}{\textbf{Rotation}}  &   \\
    &   &  \multicolumn{7}{c}{\hrulefill} &   \multicolumn{1}{c}{\hrulefill} & \multicolumn{2}{c}{\hrulefill} & \multicolumn{2}{c}{\hrulefill} &   \multicolumn{2}{c}{\hrulefill}  &   \\
$HIP$ & $(B-V)$ & $U$    & $V$    & $W$    &  $V_{Total}$ & $V_{T}^{\star}$ & $PV^{\star}$ & $\rho_{c}$ &  EW(LiI)  &  $R'_{HK}$ & Age & $ L_{X}/L_{Bol}$ &   Age     & Prot     &  Age & 
[Fe/H]\\  
    &  & (km/s)  & (km/s) & (km/s) & (km/s)      & (km/s)   & (km/s)  & (km/s)  &   (m\AA)  &  (log) & (Gyr) & (log)          &    (Gyr)  &  (days) & (Gyr) &      \\
(1) & (2) & (3) & (4) & (5) & (6) & (7) & (8) & (9) & (10) & (11) & (12) & (13) & (14) & (15) & (16) & (17) \\
\noalign{\smallskip}\hline\noalign{\smallskip}
\multicolumn{17}{c}{\textit{Probable Non-members}}\\
\noalign{\smallskip}\hline\noalign{\smallskip}
8362	&	0.79	&	-24.86	$\pm$	0.13	&	-16.48	$\pm$	0.11	&	-5.36	$\pm$	0.07	&	30.30	&	29.81	&	4.72	N	&	0.58	Y	&	7.98	$\pm$	4.22	&	-5.05	[1]	&	7.54	&	-4.96	&	0.89	&	23	[2]	&	2.53	&	0.02	\\
18267	&	0.73	&	-21.33	$\pm$	0.62	&	-24.22	$\pm$	0.41	&	-5.06	$\pm$	0.34	&	32.66	&	40.76	&	3.16	Y	&	30.95	N	&				&	-4.89	[4]	&	4.67	&	-5.17	&	1.03	&			&		&	-0.01	\\
\end{tabular}
\end{tiny}
\end{table}
\end{landscape}
}
%
%
\onltab{13}{
\begin{landscape}
\begin{table}
\centering
\begin{tiny}
\caption{Membership criteria for the Castor MG candidate stars.
(Convergence Point: 4.57 hours, -18.44 degrees; $U$ =-10.7 km/s, $V$ = -8.0 km/s, $W$ = -9.7 km/s;
Age: 200 Myr)
}
\label{castable}
\begin{tabular}{lcccccccccccccccc}
\noalign{\smallskip}\hline\noalign{\smallskip}
\hline\noalign{\smallskip}
    &   &  \multicolumn{7}{c}{\textbf{Kinematics}} &    \multicolumn{1}{c}{\textbf{Lithium}}  & \multicolumn{2}{c}{\textbf{CaII H,K}} & \multicolumn{2}{c}{\textbf{ROSAT-data}} & \multicolumn{2}{c}{\textbf{Rotation}}  &   \\
    &   &  \multicolumn{7}{c}{\hrulefill} &   \multicolumn{1}{c}{\hrulefill} & \multicolumn{2}{c}{\hrulefill} & \multicolumn{2}{c}{\hrulefill} &   \multicolumn{2}{c}{\hrulefill}  &   \\
$HIP$ & $(B-V)$ &  $U$    & $V$    & $W$    &  $V_{Total}$ & $V_{T}$ & $PV^{\star}$ & $\rho_{c}^{\star}$ &  EW(LiI)  &  $R'_{HK}$ & Age & $ L_{X}/L_{Bol}$ &   Age     & Prot     &  Age & 
[Fe/H]\\  
    &  & (km/s)  & (km/s) & (km/s) & (km/s)      & (km/s)   & (km/s)  & (km/s)  &   (m\AA)  &  (log) & (Gyr) & (log)          &    (Gyr)  &  (days) & (Gyr) &       \\
(1) & (2) & (3) & (4) & (5) & (6) & (7) & (8) & (9) & (10) & (11) & (12) & (13) & (14) & (15) & (16) & (17) \\
%
\noalign{\smallskip}\hline\noalign{\smallskip}
\multicolumn{17}{c}{\textit{Doubtful Members}}\\
\noalign{\smallskip}\hline\noalign{\smallskip}
12110	&	1.06	&	-12.52	$\pm$	0.22	&	-8.05	$\pm$	0.20	&	-11.63	$\pm$	0.13	&	18.89	&	17.47	&	0.99	Y	&	15.06	Y	&				&	-4.40	[1]	&	0.30	&	-4.81	&	1.16	&		&		&	0.06	\\
67105	&	1.09	&	-4.32	$\pm$	0.14	&	-10.06	$\pm$	0.30	&	-13.51	$\pm$	0.10	&	17.39	&	13.24	&	-7.30	N	&	-9.43	Y	&	5.15	$\pm$	5.68	&	-4.39	[1]	&	0.28	&	-5.17	&	2.15	&		&		&		\\
\end{tabular}
\end{tiny}
\end{table}
\end{landscape}
}
%
%
\onllongtabL{14}{
\label{ydtable}
\begin{landscape}
\begin{small}
\begin{longtable}{lcccccccccccc}
\caption{Properties of the stars classified as Other young discs stars}\\
\hline
\hline\noalign{\smallskip}
    &    &   \multicolumn{3}{c}{\textbf{Kinematics}} &   \multicolumn{1}{c}{\textbf{Lithium}} &
\multicolumn{2}{c}{\textbf{Ca~{\sc ii} H,K}} & \multicolumn{2}{c}{\textbf{X-ray}} & \multicolumn{2}{c}{\textbf{Rotation}} &  \\
    &    &       \multicolumn{3}{c}{\hrulefill} &  \multicolumn{1}{c}{\hrulefill} &
\multicolumn{2}{c}{\hrulefill} & \multicolumn{2}{c}{\hrulefill} & \multicolumn{2}{c}{\hrulefill} & \\
$HIP$  & $(B-V)$ & $U$   & $V$ & $W$ &  $EW$ Li~{\sc i} &
$R'_{HK}$ & $Age$ & $ L_{X}/L_{Bol}$ &  $Age$ &  $Prot$  & $Age$ & [Fe/H]  \\
   &   &  (km/s)  & (km/s) & (km/s)  & $(m\AA)$ & (log) & (Gyr) &
 (log)   & (Gyr) & (Days) & (Gyr) &  \\
 (1) & (2) & (3) & (4) & (5) & (6) & (7) & (8) & (9) & (10) & (11) & (12) & (13) \\
\endfirsthead
\multicolumn{13}{c}
{{ \tablename\ \thetable{}  Continued}} \\
\hline
\hline\noalign{\smallskip}
  &    &   \multicolumn{3}{c}{\textbf{Kinematics}} &   \multicolumn{1}{c}{\textbf{Lithium}} &
\multicolumn{2}{c}{\textbf{Ca~{\sc ii} H,K}} & \multicolumn{2}{c}{\textbf{X-ray}} & \multicolumn{2}{c}{\textbf{Rotation}} &  \\
  &    &       \multicolumn{3}{c}{\hrulefill} &  \multicolumn{1}{c}{\hrulefill} &
\multicolumn{2}{c}{\hrulefill} & \multicolumn{2}{c}{\hrulefill} & \multicolumn{2}{c}{\hrulefill} & \\
$HIP$ & $(B-V)$   &  $U$   & $V$ & $W$ &  $EW$ Li~{\sc i} &
$R'_{HK}$ & $Age$ & $ L_{X}/L_{Bol}$ &  $Age$ &  $Prot$  & $Age$ & [Fe/H]  \\
  &    &  (km/s)  & (km/s) & (km/s)  & $(m\AA)$ & (log) & (Gyr) &
 (log)   & (Gyr) & (Days) & (Gyr) &  \\
 (1) & (2) & (3) & (4) & (5) & (6) & (7) & (8) & (9) & (10) & (11) & (12) & (13)\\
\hline
\endhead
\hline  
\endfoot
\hline \hline
\endlastfoot
\noalign{\smallskip}\hline\noalign{\smallskip}
1692	&	1.31	&	-18.78	$\pm$	8.10	&	-1.08	$\pm$	3.74	&	-18.51	$\pm$	1.00	&				&			&		&		&		&			&		&		\\
3810	&	0.52	&	7.43			&	-10.88			&	-18.45			&				&	-4.92	[4]	&	5.07	&	-6.35	&	1.88	&			&		&	-0.07	\\
3998	&	1.22	&	3.40	$\pm$	0.27	&	-14.96	$\pm$	0.68	&	-23.70	$\pm$	0.68	&	21.94	$\pm$	6.34	&	-4.96	[3]	&	5.80	&	-4.53	&	1.18	&			&		&		\\
4845	&	1.10	&	5.05	$\pm$	0.32	&	-13.10	$\pm$	0.62	&	-13.53	$\pm$	0.26	&	4.35	$\pm$	5.90	&	-4.54	[1]	&	0.79	&		&		&			&		&		\\
7235	&	0.75	&	-10.89	$\pm$	0.18	&	-26.44	$\pm$	0.29	&	0.94	$\pm$	0.08	&				&	-4.72	[1]	&	2.20	&	-5.29	&	1.45	&			&		&	-0.04	\\
7918	&	0.61	&	-38.63	$\pm$	0.28	&	-31.20	$\pm$	0.26	&	-0.07	$\pm$	0.07	&	7.60	$\pm$	3.42	&	-4.99	[3]	&	6.36	&		&		&			&		&	0.06	\\
8275	&	1.02	&	-11.10	$\pm$	0.15	&	0.75	$\pm$	0.30	&	-20.62	$\pm$	0.19	&	4.53	$\pm$	5.34	&	-4.51	[1]	&	0.65	&		&		&			&		&		\\
8768	&	1.40	&	-35.88	$\pm$	0.45	&	-28.55	$\pm$	0.38	&	-1.87	$\pm$	0.22	&	26.69	$\pm$	7.20	&	-4.59	[1]	&	1.08	&	-4.52	&	1.92	&			&		&		\\
\hline
\end{longtable}
\end{small}
\end{landscape}
}


\onllongtabL{15}{
\label{otherstable}
\begin{landscape}
\begin{small}
\begin{longtable}{lcccccccccccc}
\caption{Properties of the stars non-members of moving groups}\\
\hline
\hline\noalign{\smallskip}
   &   &   \multicolumn{3}{c}{\textbf{Kinematics}} &   \multicolumn{1}{c}{\textbf{Lithium}} &
\multicolumn{2}{c}{\textbf{Ca~{\sc ii} H,K}} & \multicolumn{2}{c}{\textbf{X-ray}} & \multicolumn{2}{c}{\textbf{Rotation}} &  \\
   &   &       \multicolumn{3}{c}{\hrulefill} &  \multicolumn{1}{c}{\hrulefill} &
\multicolumn{2}{c}{\hrulefill} & \multicolumn{2}{c}{\hrulefill} & \multicolumn{2}{c}{\hrulefill} & \\
$HIP$ & $(B-V)$ &    $U$   & $V$ & $W$ &  $EW$ Li~{\sc i} &
$R'_{HK}$ & $Age$ & $ L_{X}/L_{Bol}$ &  $Age$ &  $Prot$  & $Age$ &  [Fe/H] \\
   &   &  (km/s)  & (km/s) & (km/s)  & $(m\AA)$ & (log) & (Gyr) &
 (log)   & (Gyr) & (Days) & (Gyr) &  \\
 (1) & (2) & (3) & (4) & (5) & (6) & (7) & (8) & (9) & (10) & (11) & (12) & (13) \\
\endfirsthead
\multicolumn{13}{c}
{{ \tablename\ \thetable{}  Continued }} \\
\hline
\hline\noalign{\smallskip}
  &    &   \multicolumn{3}{c}{\textbf{Kinematics}} &   \multicolumn{1}{c}{\textbf{Lithium}} &
\multicolumn{2}{c}{\textbf{Ca~{\sc ii} H,K}} & \multicolumn{2}{c}{\textbf{X-ray}} & \multicolumn{2}{c}{\textbf{Rotation}} &  \\
   &   &       \multicolumn{3}{c}{\hrulefill} & \multicolumn{1}{c}{\hrulefill}  &
\multicolumn{2}{c}{\hrulefill} & \multicolumn{2}{c}{\hrulefill} & \multicolumn{2}{c}{\hrulefill} & \\
$HIP$ & $(B-V)$ &    $U$   & $V$ & $W$ &  $EW$ Li~{\sc i} &
$R'_{HK}$ & $Age$ & $ L_{X}/L_{Bol}$ &  $Age$ &  $Prot$  & $Age$ & [Fe/H]  \\
   &   &  (km/s)  & (km/s) & (km/s)  & $(m\AA)$ & (log) & (Gyr) &
 (log)   & (Gyr) & (Days) & (Gyr) &  \\
 (1) & (2) & (3) & (4) & (5) & (6) & (7) & (8) & (9) & (10) & (11) & (12) & (13) \\
\hline
\endhead
\hline 
\endfoot
\hline \hline
\endlastfoot
\noalign{\smallskip}\hline\noalign{\smallskip}
171	&	0.65	&	-8.36			&	-76.17			&	-33.38			&				&	-4.85	[3]	&	3.96	&	-5.89	&	3.23	&			&		&	-0.52	\\
910	&	0.48	&	18.82	$\pm$	0.26	&	-12.66	$\pm$	0.23	&	-19.95	$\pm$	0.09	&	28.53	$\pm$	2.83	&	-4.79	[5]	&	3.04	&		&		&			&		&	-0.34	\\
1499	&	0.69	&	-31.47	$\pm$	0.42	&	-38.69	$\pm$	0.45	&	-1.61	$\pm$	0.17	&	9.01	$\pm$	3.75	&	-5.06	[4]	&	7.70	&		&		&			&		&	0.17	\\
1532	&	1.23	&	18.83	$\pm$	0.70	&	-26.56	$\pm$	0.85	&	1.71	$\pm$	0.38	&	42.26	$\pm$	6.47	&			&		&		&		&			&		&		\\
1598	&	0.62	&	18.02	$\pm$	0.47	&	10.64	$\pm$	0.58	&	-33.64	$\pm$	0.52	&	32.07	$\pm$	3.46	&	-4.98	[4]	&	6.21	&		&		&			&		&	-0.41	\\
1599	&	0.56	&	-71.53	$\pm$	0.12	&	-4.39	$\pm$	0.04	&	-45.35	$\pm$	0.08	&	39.59	$\pm$	3.20	&	-4.84	[7]	&	3.79	&		&		&			&		&	-0.17	\\
2941	&	0.72	&	-85.72			&	-52.37			&	-23.99			&				&	-4.90	[2]	&	4.83	&		&		&			&		&		\\
3093	&	0.85	&	39.89	$\pm$	0.31	&	-19.53	$\pm$	0.52	&	8.50	$\pm$	0.53	&	10.50	$\pm$	4.53	&	-4.99	[2]	&	6.43	&		&		&	48.0	[12]	&	8.25	&	0.14	\\
\hline
\end{longtable}
\end{small}
\end{landscape}
}


\section{Summary}
\label{summary}

 In this paper we have addressed the problem of identifying unambiguous MG members.
 Making use of a large quantity of data from
 the literature and data from our own spectroscopic observations, we were able
 to study the kinematics and age of the nearby late-type population, identifying
 a considerable group of stars that are members of moderately young (35-600 Myr)
 kinematic groups. Based on both the kinematics and different age estimates, our results allow us to
 identify new members, confirm previously suggested members of MGs, and discard  previously claimed members.

 We find that approximately $\sim$ 25\% of the nearby stars can be classified
 as members of MGs according to their kinematics, but that only 10\% have
 ages that agree with the accepted ages of the corresponding MG members. Specifically, we find
 that among the stars studied in this work, the bona fide  members for each MG
 are 14 stars (out of 29 kinematic candidates) for the Local Association, 11 (29 of  kinematic candidates)
 for the Hyades MG, six (out of 18 kinematic candidates) for the Ursa Major MG, six (out of 19 kinematic candidates)
 for IC2391, and four (out of seven kinematic candidates) for the Castor MG.  

 Some of the bona fide members identified here have not been reported before (at least to
 our knowledge), especially when considering the less-studied groups: Hyades (four new probable members),
 IC2391 (five new probable members), and Castor (three new probable members). We find discrepancies
 with previously reported lists in eight stars.   
 Additional observations are required to identify new bona fide members in each group and
 to address further investigations as suggested in Appendix~\ref{further}.


\begin{acknowledgements}
       
      We acknowledge  J. L\'opez-Santiago, I. Ribas, and J. Sanz-Forcada 
      for their valuable suggestions that contributed to improving this manuscript. 
      J.M., C.E., and B.M.  acknowledge support from
      the Spanish Ministerio de Ciencia e Innovaci\'on (MICINN),
      Plan Nacional de Astronom\'ia y Astrof\'isica, under grant
      \emph{AYA2008-01727}, and the Comunidad de Madrid project
      \emph{ASTRID S-0505/ESP/00361}. R.MA., and D.M. 
      acknowledges support from
      the Spanish Ministerio de Ciencia e Innovaci\'on (MICINN),
      Plan Nacional de Astronom\'ia y Astrof\'isica, under grant
      \emph{AYA2008-00695}, and the Comunidad de Madrid project
      \emph{AstroMadrid S2009/ESP-1496}.
      We would like to thank the staff at Calar Alto and Telescopio Nazionale Galileo for
      their assistance and help during the observing runs. 
      This research has made use of the VizieR catalogue access tool
      and the SIMBAD database, both operated at the CDS, Strasbourg, France.
      We also thank the anonymous referee for his/her valuable suggestions on how to
      improve the manuscript.

\end{acknowledgements}


\bibliographystyle{aa}
\bibliography{14948.bib}


\Online
\begin{appendix} 
\section{Tables}
\label{tables}

 Results produced in the framework of this project are published
 in electronic format only. Table~1 is also available at the CDS via
 anonymous ftp to cdsarc.u-strasbg.fr (130.79.128.5)
 or via http://cdsweb.u-strasbg.fr/cgi-bin/qcat?J/A+A/\\

 Table ~1 
 contains the following information:
 HIP number (column 1), HD number (column 2), 
 right ascension and declination (ICRSJ2000) (columns 3 and 4),
 parallax and its uncertainty (column 5),
 proper motions in right ascension and declination with their
 uncertainties (columns 6 and 7), observing run identifier (column 8),
 radial velocity  used in this work and its uncertainty (column 9), and
 radial velocities reported in KH07, NO04, NI02, and VF05 works (columns 10 to 13)
 with their uncertainties, if available.
 Column 14 contains important notes: spectroscopic binaries
 radial velocities standards, and stars 
 in chromospherically active binary systems are identified in this column.\\

 Tables ~\ref{latable} to ~\ref{castable} contain the properties of the potential candidates to 
 MG members for the different MGs studied in this work. 
 These tables give:
 HIP number (column 1), (B-V) colour (column 2), spatial-velocity components $(U,V,W)$ with their
 uncertainties (columns 3, 4 and 5), $V_{\rm{Total}}$, $V_{T}$, ${PV}$ and $\rho_{c}$
 as defined by Eggen (columns 6, 7, 8 and 9), measured  Li~{\sc i} EW (column 10),
 $R'_{\rm {H,K}}$ value and derived age (columns 11 and 12), $log(L_{\rm X}/L_{\rm Bol})$ and derived
 age (columns 13 and 14), rotational period and derived age (columns 15 and 16) and
 metallicity (column 17).
 For each Eggen's criteria, PV and $\rho_{c}$ (columns 8 and 9),
 there is label indicating if the star satisfies the criteria (label `Y') or not (label `N').\\

 Tables~14 to~15 
 are similar to the previous ones, but they show the properties
 of the stars classified as \textit{Other young disc stars}  and 
 stars not selected as possible MG members, respectively.\\

 References in Tables ~\ref{latable} to ~15 
 are indicated in parenthesis:
 (1) \cite{Raquel09}
 (2) \cite{Baliunas96}; (3) \cite{Duncan91} calculated using equations in \cite{Noyes84};
 (4) \cite{Gray03}; (5) \cite{Gray06}; (6) \cite{Hall07}; (7) \cite{1996AJ....111..439H};
 (8) \cite{Jenkins06}; (9) \cite{2005A&A...443..609S}; (10) \cite{2004ApJS..152..261W};
 (11) estimated form ROSAT-data using equation A1 in \cite{2008ApJ...687.1264M};
 (12) \cite{Noyes84}; (13) \cite{Saar97}; (14) \cite{Messina01}.\\


\section{Further applications of MGs members.}
\label{further}

 Lists of nearby MGs members constitute promising targets for a wide variety of further
 investigations. We briefly summarized some of them:

 First, we investigated whether there is a connection between the so-called
 ``solar-analogues'' \citep[e.g.][]{1997ApJ...482L..89P,2009ApJ...704L..66M,2009A&A...508L..17R}
 and MGs members. Taking as a reference the list of analogues published by \cite{Gaidos},
 we have found 25 matches between their list and our sample, where 22 out of these 25 stars,
 have been classified as bona fide MG members.   
 Another three stars, HIP 29525, HIP 80337, and HIP 116613, also candidates for MGs, satisfy
 Gaidos' criteria for being considered as solar analogues. These stars are listed in Table ~\ref{solar}.
 These ``young-suns'' are essential to study the history and formation of our own Solar System, 
 indeed three of them, namely
 HIP 15457, HIP 42438, and HIP 64394, are included in the ambitious 
 project \textit{The Sun in Time} aimed at reconstructing the spectral irradiance evolution
 of the Sun \citep[e.g][]{2005ApJ...622..680R}.
 
 As we have shown in Section~\ref{secdebris}, debris discs are linked to stars in MGs.
 It is therefore natural to check if there is a similar relation between stars with known
 planets and MGs. Nineteen stars of our sample have detected planets
 \footnote{The Extrasolar Planets Encyclopedia,\\ http://www.obspm.fr/encycl/es-encycl.html}, and
 seven of them are MGs candidates: HIP 21482 (Hyades MG, but  it is doubtful member and in addition the
 planet is not confirmed); HIP 43587 (Hyades MG, but it is a doubtful member), HIP 71395
 (Ursa Major MG, probable member); HIP 80337 (Ursa Major, doubtful member); HIP 95319 (IC2391 MG,
 but doubtful member and the planet is not confirmed);
 HIP 49669 and HIP 53721 (young disc stars according to their kinematics, but their
 calcium ages suggest that they are old stars).

 Other applications are related to activity studies, i.e., flux-flux and rotation-activity-age
 relationships \citep[e.g][]{Raquel09}, or search programmes to detect stellar and substellar
 companions \citep[e.g][]{2007A&A...463..707H}. Finally, we point out that an important fraction of
 the stars analysed in this paper will be observed in the framework of the DUNES
 (DUst around NEarby Stars) programme, an approved Herschel
 Open Time Key Project with the aim of detecting cool faint dusty discs,
 at flux levels as low as the Solar EKB
 \citep{Maldonado10,2010arXiv1005.3151E}.


\begin{table}
\centering
\begin{tiny}
\caption{Solar analogues and their ascription to MGs.Label `Y' indicates
probable members, `?' doubtful members, and `N' probable non-members,
respectively.}
\label{solar}
\begin{tabular}{lcclcc}
\hline\noalign{\smallskip}
HIP     & MG            & Membership    & HIP           &  MG          &  Membership    \\
\noalign{\smallskip}\hline\noalign{\smallskip}
544     &       LA      &       Y       &       46843   &       LA      &       Y       \\
1803    &       HS      &       Y       &       54745   &       LA      &       Y       \\
5944    &       UMa     &       Y       &       63742   &       LA      &       Y       \\
7576    &       LA      &       Y       &       65515   &       LA      &       Y       \\
8362    &       IC2     &       N       &       71743   &       IC2     &       Y       \\
8486    &       UMa     &       Y       &       72567   &               &               \\
15457   &       YD      &       Y       &       74702   &       UMa     &       N       \\
22263   &       IC2     &       Y       &       77408   &       LA      &       Y       \\
26779   &       LA      &       Y       &       80337   &       UMa     &       ?       \\
29525   &       YD      &       Y       &       82588   &               &               \\
29568   &       IC2     &       Y       &       94346   &       HS      &       Y       \\
42074   &       HS      &       Y       &       107350  &       LA      &       Y       \\
42333   &       HS      &       Y       &       115331  &               &               \\
42438   &       UMa     &       Y       &       116613  &       HS      &       Y       \\
\hline
\multicolumn{6}{l}{LA: Local Association; HS: Hyades; UMa: Ursa Major}\\
\multicolumn{6}{l}{IC2: IC2391; Cas: Castor}\\
\noalign{\smallskip}\hline\noalign{\smallskip}
\end{tabular}
\end{tiny}
\end{table}

\end{appendix}

\end{document}